\documentclass[aps,prb,twocolumn,superscriptaddress]{revtex4-2}
\usepackage{xcolor}

\usepackage{amsmath}
\usepackage{amssymb}
\usepackage{commath}
\usepackage{bm}
\usepackage{times}
\usepackage{physics}
\usepackage{CJKutf8}

\usepackage[colorlinks = true,
            linkcolor = blue,
            urlcolor  = blue,
            citecolor = blue,
            anchorcolor = blue]{hyperref}
\usepackage{cleveref}
\usepackage{graphicx}

\begin{document}
\crefname{equation}{Eq.}{Eqs.}
\crefname{figure}{Fig.}{Figs.}
\crefname{section}{Sec.}{Secs.}
\crefname{appendix}{Appendix}{Appendices}

\definecolor{purple}{rgb}{.6,.1,.6}
\definecolor{darkgreen}{rgb}{.1,.6,.1}
\newcommand{\ZYW}[1]{\textcolor{purple}{ZYW:#1}}
\newcommand{\ZYWe}[1]{\textcolor{purple}{#1}}
\newcommand{\EC}[1]{\textcolor{blue}{[EC: #1]}}
\newcommand{\DM}[1]{\textcolor{darkgreen}{DM: #1}}
\newcommand{\todo}[1]{\textcolor{red}{#1}}

\title{Noise-induced suppression of MPO truncation errors\\ in noisy random circuits and Lindbladian dynamics}

\author{Zhi-Yuan Wei \begin{CJK}{UTF8}{gbsn}(魏志远)\end{CJK}}
\email{zywei@umd.edu}
\affiliation{Joint Center for Quantum Information and Computer Science,
NIST/University of Maryland, College Park, Maryland 20742, USA}
\affiliation{Joint Quantum Institute,
NIST/University of Maryland, College Park, Maryland 20742, USA}

\author{Joel Rajakumar}
\affiliation{Joint Center for Quantum Information and Computer Science,
NIST/University of Maryland, College Park, Maryland 20742, USA}

\author{Jon Nelson}
\affiliation{Joint Center for Quantum Information and Computer Science,
NIST/University of Maryland, College Park, Maryland 20742, USA}

\author{Jens~Eisert}
\affiliation{Dahlem Center for Complex Quantum Systems, Freie Universit\"{a}t Berlin, 
14195 Berlin, Germany}
\affiliation{Helmholtz-Zentrum Berlin f{\"u}r Materialien und Energie, 14109 Berlin, Germany}
\affiliation{Fraunhofer Heinrich Hertz Institute, 10587 Berlin, Germany}

\author{Daniel~Malz}
\affiliation{Department of Physics, University of Basel, CH-4056 Basel, Switzerland}
\affiliation{Department of Mathematical Sciences, University of Copenhagen, 2100 Copenhagen, Denmark}

\author{Michael J. Gullans}
\affiliation{Joint Center for Quantum Information and Computer Science,
NIST/University of Maryland, College Park, Maryland 20742, USA}
\affiliation{National Institute of Standards and Technology, Gaithersburg, MD 20899, USA.}

\author{Alexey V. Gorshkov}
\affiliation{Joint Center for Quantum Information and Computer Science,
NIST/University of Maryland, College Park, Maryland 20742, USA}
\affiliation{Joint Quantum Institute,
NIST/University of Maryland, College Park, Maryland 20742, USA}
\affiliation{National Institute of Standards and Technology, Gaithersburg, MD 20899, USA.}

\date{\today}

\begin{abstract}
We study how \emph{matrix product operator} (MPO) truncation errors evolve when simulating two setups: (1) one-dimensional (1D) Haar-random circuits under either depolarizing noise or amplitude-damping noise, and (2) 1D Lindbladian dynamics of a non-integrable quantum Ising model under either depolarizing or amplitude-damping noise. We first show that the average purity of the system density matrix relaxes to a steady value on a timescale that scales inversely with the noise rate. We find that truncation errors generated at earlier steps are approximately exponentially contracted by the subsequent noisy evolution, both with system size $N$ and with the evolution depth $t$, while the cumulative error results from the competition between this contraction and newly injected compression errors. We complement the local-circuit numerics with a solvable global random-unitary model in which the same channel-dependent factor governs the mean-squared contraction of traceless errors and transient mean purity. Its exact finite-$N$ results capture the weak-noise contraction rates and the non-unital stationary-purity trend. Combining the resulting $L_2$ error characterization with the
observed late-time $L_1/L_2$ conversion yields an empirical estimate
of the $L_1$ truncation error that can be exponentially tighter in
$N$ than the generic rank-based bound, particularly for non-unital
noise. Together, these results provide empirical evidence that MPO simulation algorithms may efficiently sample from the output of 1D noisy random circuits [setup (1)] with an arbitrary number of circuit layers, and from the steady state of 1D Lindbladian dynamics [setup (2)].
\end{abstract}

\maketitle
\section{Introduction}

Open quantum systems are ubiquitous, both in nature and in engineered quantum devices~\cite{breuerbook02a}. In many settings, noise qualitatively reshapes the dynamics: it can drive decoherence and wash out quantum effects~\cite{aharonov1996limitations}, but when properly engineered it can also induce novel phases of matter~\cite{coser2019classification,de2022symmetry,PhysRevX.13.031016} or prepare states that are useful for quantum information processing~\cite{Kraus2008,Diehl2008b,Verstraete2009,doi:10.1126/science.adh9932}. Efficiently simulating such open quantum dynamics is therefore of central importance~\cite{RevModPhys.93.015008}. Efficient simulation enables a deeper understanding of these emerging phenomena and provides practical benchmarks for assessing how noise impacts near-term quantum hardware, ultimately clarifying when—and for what tasks—noisy quantum devices can offer a computational advantage~\cite{hangleiter2023computational}.
Noise can also facilitate classical simulation by
limiting the effective depth of noisy quantum
circuits. Previous work has established
classical-simulability results for noisy optimization
algorithms~\cite{stilck2021limitations} and efficient
estimation of local observables in random circuits
under non-unitary noise~\cite{mele2026noise}.

A widely used approach for simulating open quantum systems is to represent the density matrix as a \emph{matrix product operator}  (MPO)~\cite{verstraete2004matrix,pirvu2010matrix,Cirac2017a,Werner2016}. 
MPOs have been used in multiple contexts, such as simulating thermal states~\cite{verstraete2004matrix,Alhambra2021,Kuwahara2021,PhysRevX.4.031019}, non-equilibrium steady states~\cite{Prosen_2009,Cui2015}, Lindbladian dynamics~\cite{verstraete2004matrix,RevModPhys.93.015008}, non-Markovian dynamics~\cite{Strathearn2018,PRXQuantum.3.010321}, and operator dynamics for quantum chaos~\cite{xu2020accessing}. They have also found broad use in quantum-information settings, such as noisy random-circuit evolution~\cite{Noh2020,zhang2022entanglement,wei2025measurement}, density matrix tomography~\cite{3vnw-yjd5,Baumgratz2013b,Baumgratz2013}, quantum error mitigation~\cite{PRXQuantum.3.040313}, and noisy boson sampling~\cite{PhysRevA.108.052604,PhysRevA.104.022407}.

However, despite its wide applicability, MPO simulation often lacks clear theoretical error guarantees. In particular, the standard MPO truncation scheme [cf.~\cref{trunc_err_sec}] controls the truncation error in the $L_2$ norm. Converting this into an $L_1$ bound can introduce a factor exponential in system size. For density matrices, half the $L_1$ distance is the trace distance, which quantifies optimal state distinguishability and upper bounds the \emph{total variation distance}  (TVD) of measurement outcomes~\cite{Nielsen2000}. This gap limits the use of MPO algorithms in settings where an $L_1$ guarantee is required (see \cref{l1_naive_bound} for details). Therefore, although operator entanglement (an indicator of MPO efficiency) obeys an area law in noisy random circuits~\cite{Noh2020,wei2025measurement}, noisy boson sampling processes~\cite{PhysRevA.108.052604,PhysRevA.104.022407}, and 1D Lindbladian dynamics~\cite{PhysRevA.108.012616,PhysRevLett.129.170401}, it remains unclear whether these MPO simulations can sample from the output distributions within small TVD. Although one can control the $L_1$ error of an MPO representation using a renormalization-group–style truncation~\cite{PRXQuantum.1.010304}, that approach complicates the algorithm and requires computing the entanglement of purification, which is generally intractable for many-body systems.

The existing bound on the $L_1$ truncation error in terms of the $L_2$ truncation error [cf.~\cref{l1_naive_bound}] is generally expected to be loose. Moreover, it neglects a potentially beneficial feature of open-system dynamics, namely, that the dynamics itself may suppress previously generated errors. Noise can suppress distinctions between density-matrix trajectories~\cite{PhysRevA.109.022218,aharonov1996limitations,frigerio1977quantum}, thereby reducing the distance between the true density matrix and its MPO approximation. This raises a basic question: \textit{How large are the errors incurred by MPO algorithms when simulating open quantum dynamics, and how do these errors evolve under the same dynamics?} Understanding such ``error dynamics'' would deepen our understanding of MPO-based simulation and could lead to tighter truncation-error estimates, thereby placing the empirical success of MPO algorithms across a range of settings on firmer theoretical footing.

In this work, we study how MPO truncation errors evolve in 1D noisy Haar-random circuits and in 1D Lindbladian dynamics, under either (unital) depolarizing noise or (non-unital) amplitude-damping noise. In \cref{sec_setup}, we introduce the two setups and the corresponding MPO simulation algorithms, and review the existing error bound, highlighting why the existing bound fails to provide meaningful guarantees on the $L_1$ error. We also provide  intuition for why noisy quantum dynamics may help reduce simulation errors.

In \cref{circ_num_sec}, we present results for 1D noisy random circuits under either depolarizing or amplitude-damping noise. We first show that the $L_2$ norm of the density matrix decays exponentially with system size $N$ at fixed positive evolution times, and relaxes to its steady-state value on a timescale $T_s\sim 1/p$, where $p$ is the noise rate. We then analyze the evolution of the $L_2$ truncation error and identify a \emph{noise-induced suppression} effect, whereby the noisy dynamics exponentially suppresses the $L_2$ error in both $N$ and the number of circuit layers $t$. Combining these observations, we derive an empirical estimate for the total $L_2$ error that is proportional to the $L_2$ norm and hence exponentially small in $N$, and we numerically verify that this empirical estimate agrees well with the observed $L_2$ error. 

Moreover, once the evolution reaches the steady state, we find that the ratios among the $L_1$ error, the $L_2$ error, and the $L_2$ norm appear to approach values that are independent of system size over the accessible sizes. We further relate this behavior to the normalized singular-value spectra of two matrices: the steady-state density matrix $\rho_{\infty}$ and the error matrix $\rho_{\infty}-\sigma_{\infty}$, where $\sigma_{\infty}$ is the normalized MPO approximation in the stationary late-time regime. This enables us to obtain an empirical \(L_1\)-error estimate that can substantially improve on the generic rank-based bound and is exponentially tighter in the non-unital amplitude-damping case. Taken together, our results suggest an MPO-based simulation strategy that may enable efficient sampling from 1D noisy random circuits of arbitrary depth with small TVD.

In \cref{haar_toy_sec}, we develop a solvable global random-unitary model for the observed $L_2$ suppression.  Haar conjugation makes the quadratic moment of a traceless error isotropic, reducing its mean-squared contraction to the Frobenius action of the noise channel on the traceless operator space.  The resulting exact finite-$N$ formulas for depolarizing and amplitude-damping noise predict exponential-in-$N$ error contraction.  The same contraction factor governs the transient mean purity, explaining within the solvable model why the state $L_2$ norm and a pre-existing traceless error have the same decay exponent.  The weak-noise contraction rates agree well with the local brickwall data at the smallest noise rates studied, while the amplitude-damping solution also reproduces the nontrivial stationary-purity trend and inverse-noise crossover scale.

In \cref{lind_num_sec}, we extend our study to 1D Lindbladian dynamics under either depolarizing or amplitude-damping noise and perform an analysis parallel to \cref{circ_num_sec}. Although Lindbladian dynamics differ in important ways from noisy random circuits, we again observe the same qualitative behavior of MPO truncation errors, including exponential suppression of the $L_2$ norm of the density matrix, noise-induced suppression of truncation errors, and late-time size collapse of the $L_1/L_2$ conversion factor $\Lambda_t$. This likewise yields a substantially tighter empirical characterization of the \(L_1\) error, with an exponential improvement over the generic rank-based bound in the non-unital 
cases studied, and suggests that standard MPO methods may serve as efficient algorithms for sampling from the steady state of the 1D Lindbladian dynamics studied in our work with small TVD.
Finally, we provide an outlook in \cref{sec_discuss}.

\section{Setups and their MPO simulation}
\label{sec_setup}
In this section, we introduce the two setups considered in this work and formulate the corresponding MPO algorithms used to simulate them [\cref{circ_set,lind_set}]. Then, in \cref{trunc_err_sec}, we discuss a known bound on the single-step $L_2$ truncation error, along with a loose bound on the $L_1$ error. Finally, in \cref{sec_pre_intu}, we provide intuition for why noisy evolution could reduce the errors incurred by the MPO simulation algorithm.

\subsection{One-dimensional noisy random-circuit evolution}
\label{circ_set}
\subsubsection{Setup}
\begin{figure}[h!]
	\centering
\includegraphics[width=0.48\textwidth]{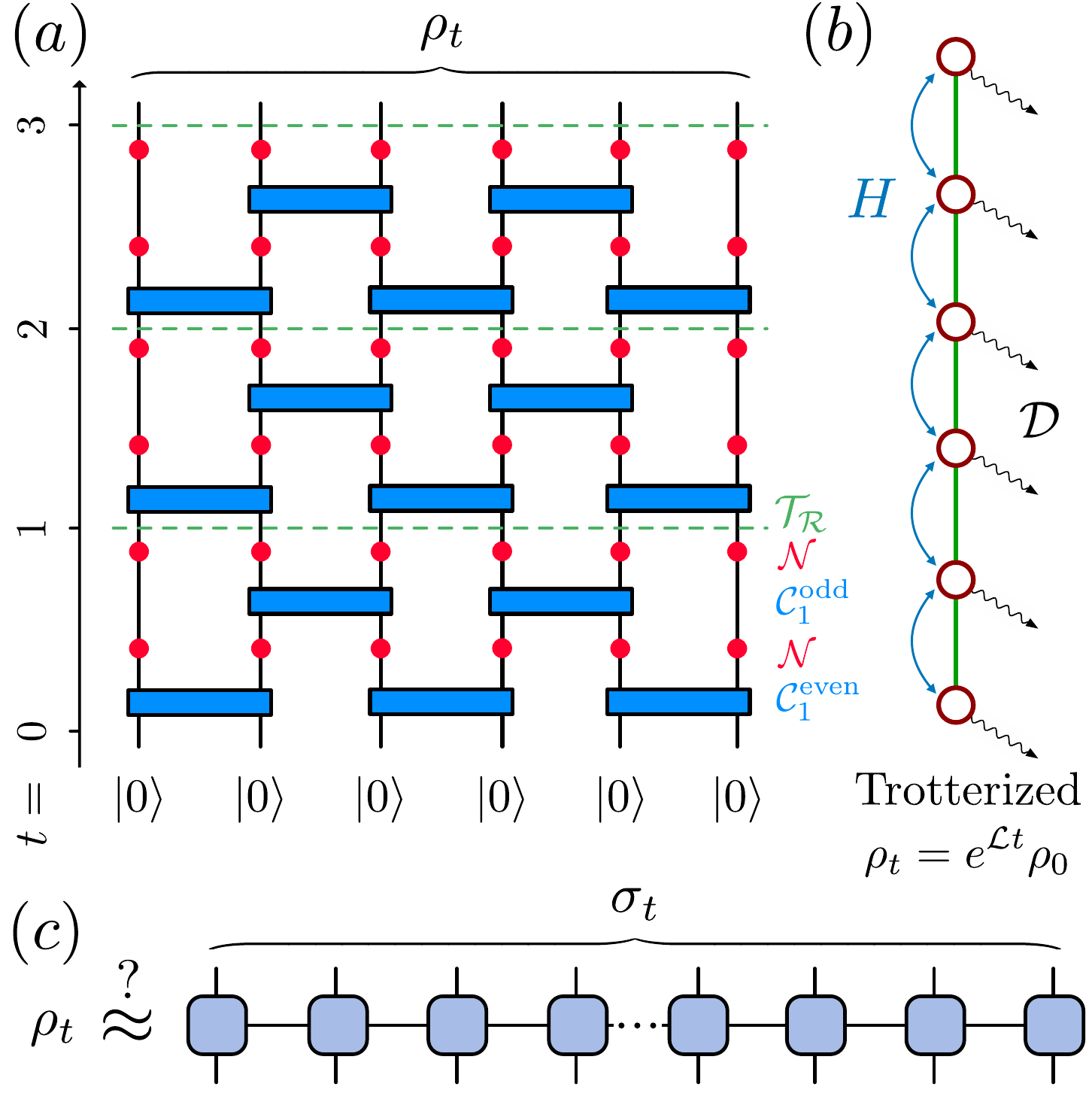}
        \caption{\textbf{Setups}. (a) The 1D brickwall circuit of $N$ qubits, where the gates are denoted by blue boxes. After each gate layer, single-qubit noise (depolarizing or amplitude-damping, denoted as red circles) is applied. The locations of the gate layers ${\cal C}^{\rm even}_1$ and ${\cal C}^{\rm odd}_1$, as well as the noise layer ${\cal N}$, are marked. The exact circuit evolution produces a density matrix $\rho_t$.  The MPO algorithm simulates the same circuit dynamics, with a truncation followed by normalization (denoted together as ${\cal T_R}$) applied after each full layer of unitary-noise evolution, as indicated by the green dashed lines, resulting in a unit-trace MPO approximation $\sigma_t$.  (b) Illustration of 1D Trotterized Lindbladian dynamics of $N$ qubits (hollow circles), consisting of evolution under a local Hamiltonian $H$ (curvy blue arrows) and a dissipator $\cal D$ describing single-qubit noise (depolarizing or amplitude-damping, denoted as wavy black arrows). (c) The corresponding MPO algorithm simulates the dynamics in panels (a,b) approximately, producing an approximation $\sigma_t$ to the density matrix $\rho_t$. We are interested in how the distance between $\rho_t$ and $\sigma_t$ evolves over time.}
        \label{fig1}
\end{figure}

We consider a 1D system of $N$ qubits with basis $|0\rangle, |1\rangle$, initialized in the state $\rho_{0}=|0,\dots, 0\rangle\langle0,\dots, 0|$, and evolved under Haar-random brickwall circuits [\cref{fig1}(a)]. The system is subject to single-qubit noise applied after each commuting gate layer. We define one full circuit layer as an even layer followed by an odd layer, so that each full layer entangles the entire lattice once. We will consider two types of noise: depolarizing noise and  amplitude-damping noise. Depolarizing noise is a paradigmatic example of unital noise, which is widely used to model decoherence as well as coherent noise after Pauli twirling in quantum devices~\cite{Arute2019a,10.1145/3564246.3585234,emerson2005scalable}. On the other hand, amplitude damping is a paradigmatic example of non-unital noise, modeling $T_1$ relaxation in quantum devices~\cite{carroll2022dynamics,PhysRevX.13.041022}, and can in principle enable quantum refrigeration~\cite{ben2013quantum}.

For any matrix input $A$, a quantum channel $\Phi$ is characterized by its Kraus operators $\{K_\mu \}$ as $\Phi(A)=\sum_\mu K_\mu A K_\mu^{\dagger}$. 
For a single-qubit depolarizing noise channel acting on the $i$-th qubit with rate $0\le p_{\rm dep}\le 1$, the Kraus operators are
\begin{equation} \label{kraus_dep}
K_{0,i}^{\rm dep}=\sqrt{1-p_{\rm dep}}\mathbb I_i, \quad K_{\mu \in\{x,y,z\},i}^{\rm dep}=\sqrt{\frac{p_{\rm dep}}{3}}\tau_i^\mu,
\end{equation}
where $\mathbb I_i$ and $\tau_i^\mu$, $\mu\in\{x,y,z\}$ are the identity and the Pauli operators acting on the $i$-th qubit. In the case of single-qubit amplitude-damping noise with rate $0\le p_{\rm damp}\le 1$, the Kraus operators are
\begin{equation} \label{kraus_damp}
K_0^{\rm damp}=\left(\begin{array}{cc}
1 & 0 \\
0 & \sqrt{1-p_{\rm damp}}
\end{array}\right), \quad K_1^{\rm damp}=\left(\begin{array}{cc}
0 & \sqrt{p_{\rm damp}} \\
0 & 0
\end{array}\right) .
\end{equation}

The circuit evolution in the $t$-th layer consists of an even and an odd commuting gate layer, represented by the unitary channels ${\cal C}^{\rm even}_{t}$ and ${\cal C}^{\rm odd}_{t}$, respectively. The noise channel ${\cal N}$ is the same for every layer, as illustrated in \cref{fig1}(a). Thus, in this layer, the system evolves under the composite channel 
\begin{equation} 
{\cal K}^{\cal C}_{t}={\cal N}\circ {\cal C}^{\rm odd}_{t} \circ{\cal N}\circ {\cal C}^{\rm even}_{t},	
\end{equation}
and the evolution of the density matrix $\{\rho_t\}$ of the system is
\begin{equation} \label{noi_evo}
	\rho_{t}={\cal K}^{\cal C}_{t}(\rho_{t-1}).
\end{equation}

\subsubsection{MPO simulation of noisy circuits}
\label{sec_MPO_algo}

We now turn to discussing the actual MPO simulation of noisy quantum circuits.
The MPO approach simulates the system evolution [cf.~\cref{noi_evo}] by representing both the initial state and the quantum channels as MPOs, and by applying the channel MPO to the state MPO to propagate the state. During the evolution, the state MPO is truncated (the truncation operation denoted by ${\cal T}$) after each layer of the composite channel ${\cal K}_{t}^{\cal C}$ consisting of two unitary layers, each followed by a noise layer. After each truncation, we also renormalize the MPO to preserve the trace; we denote this operation by ${\cal R}$. We define the combined operation as ${\cal T_R} = {\cal R}\circ{\cal T}$ [cf.~\cref{trunc_err_sec}]. With this notation, the MPO approximation of the density matrix, $\sigma_t$, evolves as [cf.~\cref{fig1}(a)]
\begin{equation} \label{mpo_evo}
	\sigma_{t}=({\cal T_R}\circ{\cal K}_{t}^{\cal C})(\sigma_{t-1}),
\end{equation}
with initial state $\sigma_{0}=\rho_{0}$.

Trace normalization alone does not guarantee positivity or Hermiticity after an unconstrained MPO truncation. We therefore use $\sigma_t$ to denote a unit-trace MPO approximation, without assuming that it is a density matrix. The norm estimates below remain meaningful for such operators. Where a measurement interpretation is needed, Hermiticity can be imposed by replacing $\sigma_t$ with $(\sigma_t+\sigma_t^\dagger)/2$; this does not increase either the $L_1$ or $L_2$ error relative to Hermitian $\rho_t$, and increases the MPO bond dimension by at most a factor of two. Positivity and sampling are discussed separately in \cref{mpo_circ_samp_algo}.

\subsection{One-dimensional  Lindbladian evolution}
\label{lind_set}
\subsubsection{Setup and the Trotterization}

\textit{Formulation.} Besides the 1D noisy random circuit evolution in \cref{circ_set}, we also consider the same 1D $N$-qubit system, initialized in the state $\rho_{0}^{\cal L}=|+\ldots +\rangle\langle +\ldots +|$, with $|+\rangle = (|0\rangle + |1\rangle)/\sqrt{2}$, evolving under the Lindblad dynamics [cf.~\cref{fig1}(b)]
\begin{equation}\label{lind_equ}
\dot\rho = \mathcal L(\rho)=\mathcal L_H(\rho)+\mathcal D(\rho),
\quad \mathcal L_H(\rho)=-i[H,\rho],
\end{equation}
where $H$ is the Hamiltonian and the dissipator is
\begin{equation}\label{diss_equ}
\mathcal D(\rho)=\sum_\mu\sum_{i=1}^N \Bigl(L_{i,\mu}\rho L_{i,\mu}^\dagger
-\frac12\{L_{i,\mu}^\dagger L_{i,\mu},\rho\}\Bigr),
\end{equation}
with ${L_{i,\mu}}$ denoting the single-qubit jump operators (with the type indexed by $\mu$). In particular, in this work, we focus on the non-integrable quantum Ising model with both transverse and longitudinal fields,
\begin{equation}\label{hami_equ}
H
= -J\sum_{i=1}^{N-1}\tau_i^z\tau_{i+1}^z
- g\sum_{i=1}^{N}\tau_i^x
- h\sum_{i=1}^{N}\tau_i^z,
\end{equation}
where we set $J=1$ to fix the units of energy and time (recall that $\tau_i^\mu$, with $\mu\in\{x,y,z\}$, are the Pauli operators). We also consider continuous depolarizing noise (with rate $\kappa_{\rm dep}$) and amplitude-damping noise (with rate $\kappa_{\rm damp}$), with the corresponding jump operators
\begin{align}\label{diss_form}
&\textrm{Depolarizing:}\qquad
L_{i,\mu \in\{x,y,z\}}^{\rm dep}=\sqrt{\frac{\kappa_{\rm dep}}{3}}\,\tau_i^{\mu}, \nonumber\\
&\textrm{amplitude-damping:}\qquad
L_{i}^{\rm damp}=\sqrt{\kappa_{\rm damp}}|0\rangle_i \langle 1|.
\end{align}

\textit{Trotterized evolution.} Continuous Lindbladian dynamics are typically Trotterized to enable digital quantum simulation or classical simulation. In the following, we briefly describe the Trotterization scheme used in this work; further details are provided in Appendix~\ref{app:lindblad_trotter}.

We consider a second-order (Strang) splitting between coherent and dissipative parts. The density matrix $\rho_{t}$ produced by this Trotterized evolution satisfies
\begin{equation} \label{lind_evo}
\rho_{t}=
{\cal K}^{\cal L}_{\Delta t}\bigl(\rho_{t-\Delta t}\bigr),
\end{equation}
with the composite channel 
\begin{equation} \label{lind_tro_channel}
{\cal K}^{\cal L}_{\Delta t}=	\bigl(\mathcal E_{\Delta t/2}^{\otimes N}\bigr)\circ
\mathcal U_{\Delta t}\circ
\bigl(\mathcal E_{\Delta t/2}^{\otimes N}\bigr).
\end{equation}
Here $\mathcal U_{\Delta t}$ is a second-order brickwall Suzuki–Trotter approximation of the unitary channel generated by $H$, and $\mathcal E_{\Delta t}^{\otimes N}$ is the channel generated by the dissipator ${\cal D}$ (amplitude-damping or depolarizing). Each single-qubit channel $\mathcal E_{\Delta t}$ uses the same set of Kraus operators as in \cref{kraus_dep,kraus_damp}, parameterized as
\begin{align}
    	p_{\rm dep}(\Delta t)&=\frac{3}{4}(1-e^{-4\kappa_{\rm dep} \Delta t/3}), \\
        p_{\rm damp}(\Delta t)&=1-e^{-\kappa_{\rm damp} \Delta t}.
\end{align}
The parametrization is chosen so that $\mathcal E_{\Delta t}$ is the finite-time
single-site channel generated by the dissipator in \cref{diss_form}. In particular, both reduce to $p_{\alpha}=\kappa_{\alpha}\Delta t+O(\Delta t^2)$, with $\alpha=\mathrm{dep},\mathrm{damp}$.

\subsubsection{MPO simulation of Trotterized Lindbladian dynamics}
\label{mpo_lind_algo_sec}
Similar to the noisy-circuit case [cf.~\cref{sec_MPO_algo}], to simulate the Trotterized Lindbladian evolution [cf.~\cref{lind_evo}] with MPO, we represent the initial state and the quantum channels as MPOs. After each time step $\Delta t$, we truncate and renormalize the state MPO. The MPO approximation $\sigma_t$ to the density matrix $\rho_t$ therefore evolves as
\begin{equation} 
\sigma_{t}=({\cal T_R}\circ{\cal K}^{\cal L}_{\Delta t})(\sigma_{t-\Delta t}),
\end{equation}
with initial condition $\sigma_{0}=\rho_{0}^{\cal L}$.

Note that in this work we focus on the truncation error, characterized by the distance between the density matrix $\rho_t$ and its MPO approximation $\sigma_t$ in the Trotterized Lindbladian evolution. At the same time, Trotterization also introduces a Trotter error between $\rho_t$ and the true density matrix under the continuous Lindbladian evolution in \cref{lind_equ}. Thus, by the triangle inequality, the total distance between the true density matrix and $\sigma_t$ produced by the MPO simulation can be upper bounded by the sum of the Trotter error and the truncation error considered here. Since the Trotter error in the $L_1$ norm can be systematically suppressed by increasing the Trotterization order or reducing the Trotter step $\Delta t$~\cite{PhysRevLett.107.120501}, we focus here on characterizing the truncation error introduced by the MPO approximation.

\subsection{MPO truncation and error bounds}
\label{trunc_err_sec}
In this work, we are interested in the difference $\rho_t-\sigma_t$ between the density matrix $\rho_t$ and its MPO approximation $\sigma_t$ for both 1D noisy circuit evolution and Lindbladian dynamics [cf.~\cref{circ_set,lind_set}]. This difference evolves under the combined effects of the system’s physical dynamics and the MPO truncation performed after each evolution step. To set the stage, we briefly review some standard results on matrix norms, along with known bounds on MPO truncation errors.

\subsubsection{Matrix norms}
\label{norm_sec}

Throughout, $L_1$ and $L_2$ mean the Schatten $1$- and $2$-norms, respectively, rather than entrywise matrix norms; in particular, $L_2$ is the Hilbert--Schmidt (Frobenius) norm, not the spectral norm. The $L_2$ norm is natural for MPO compression because vectorization is an isometry into a Hilbert space, and a Schmidt truncation minimizes the squared error across a single bipartition. Purities, overlaps, and squared compression residuals can also be evaluated by MPO contractions. These computational advantages do not by themselves give an operational error guarantee.
The $L_2$ norm of a matrix $A$ is defined as $\|A\|_2 := \sqrt{\operatorname{Tr}\left(A^{\dagger} A\right)}$. For a density matrix $\rho$, the $L_2$ norm is the square root of the purity, as $\|\rho\|_2 = \sqrt{{\rm Tr}[\rho^2]}$.

For a matrix $A$, the $L_1$ norm is $\|A\|_1:=\operatorname{Tr}\!\left(\sqrt{A^{\dagger} A}\right)$. The following relation holds between the $L_1$ and $L_2$ norms,
\begin{equation} \label{L12_relation}
\|A\|_2 \leq\|A\|_1 \leq \sqrt{\operatorname{rank}(A)}\|A\|_2.
\end{equation}

\subsubsection{The single-step truncation-and-normalization error}

In the MPO algorithms considered in this work, each compression step
consists of a truncation $\mathcal{T}$ followed by trace normalization $\mathcal R$. We denote the
combined operation by
\begin{equation}
    \mathcal{T}_{\mathcal R}
    =
    \mathcal R\circ\mathcal T,
    \qquad
    \mathcal{T}_{\mathcal R}(\rho)
    =
    \frac{\mathcal T(\rho)}
    {\Tr[\mathcal T(\rho)]}.
\end{equation}
We now characterize the single-step $L_2$ error introduced by this
combined operation.

Consider an $N$-qubit density matrix $\rho$ represented as an MPO
[cf.~\cref{fig1}(c)]. To truncate the bond between the $k$-th and
$(k+1)$-th qubits, we first vectorize the MPO as an unnormalized MPS
using the Hilbert--Schmidt isomorphism
\begin{equation}
A=\sum_{i,j}A_{i,j}|i\rangle\langle j|
\quad\mapsto\quad
|A\rangle=\sum_{i,j}A_{i,j}|i\rangle\otimes|j\rangle.
\end{equation}
We compute the Schmidt values $\{\lambda_{k,j}\}$ across the $k$-th
bond, ordered as
$\lambda_{k,1}\geq\lambda_{k,2}\geq\cdots$, and retain only the largest
$D_k$ values. The relative discarded weight is
\begin{equation}
\delta_k(D_k)
=
\frac{\sum_{j>D_k}\lambda_{k,j}^{2}}
{\sum_j\lambda_{k,j}^{2}}.
\end{equation}
Given a threshold $\delta_{\rm err}$, the bond dimension $D_k$ is chosen
as the smallest value satisfying
\begin{equation}
    \delta_k(D_k)\leq\delta_{\rm err},
    \qquad k=1,\ldots,N-1.
\end{equation}
To bound the error after both truncation and normalization, let
\begin{equation}
    \widetilde{\rho}=\mathcal T(\rho),
    \qquad
    \Xi=\rho-\widetilde{\rho},
    \qquad
    \epsilon=\Tr \Xi
    =1-\Tr\widetilde{\rho}.
\end{equation}
The standard MPS truncation estimate gives~\cite{Verstraete2006c,baumgratz2014efficient}
\begin{equation}
    \|\Xi\|_2
    \leq
    \sqrt{2\sum_{k=1}^{N-1}\delta_k(D_k)}\,\|\rho\|_2
    \leq
    \sqrt{2(N-1)\delta_{\rm err}}\,\|\rho\|_2.
    \label{eq:intermediate_truncation_bound}
\end{equation}
Here, this estimate is used only as an intermediate step; the output of
the MPO compression procedure is the normalized operator $\mathcal T_{\mathcal R}(\rho)$. Using
\begin{equation}
    \rho-\mathcal T_{\mathcal R}(\rho)
    =
    \rho-\frac{\rho-\Xi}{1-\epsilon} 
    =
    \frac{\Xi-\epsilon\rho}{1-\epsilon},
\end{equation}
the triangle inequality and
$|1-\epsilon|\geq 1-|\epsilon|$ imply, provided $|\epsilon|<1$,
\begin{equation} 
    \|\mathcal T_{\mathcal R}(\rho)-\rho\|_2
    \leq
    \frac{
    \sqrt{2(N-1)\delta_{\rm err}}+|\epsilon|
    }{
    1-|\epsilon|
    }\,
    \|\rho\|_2.
    \label{err_bound_normalize_formal}
\end{equation}
Equation~\eqref{err_bound_normalize_formal} is the formal single-step
error bound for the combined truncation-and-normalization operation. It shows
that normalization introduces an additional contribution controlled by
the trace deviation $\epsilon$. A sharper a posteriori bound can be obtained by separating
the traceless component of the compression residual.
We give this refinement, together with an alternative
affine trace correction, in Appendix~\ref{app:normalization}.

\cref{normalize_effect_fig}(a,b) show the representative behavior of the trace deviation $\epsilon$ as a function of the system size $N$ and the MPO truncation threshold $\delta_{\rm err}$ for noisy random-circuit evolution under depolarizing noise ($p_{\rm dep}=0.02$, $t=1$) and amplitude-damping noise ($p_{\rm damp}=0.1$, $t=1$). We find empirically that the trace deviation $\epsilon>0$, and scales as
\begin{equation}
    \epsilon
    \lesssim
    cN\sqrt{\delta_{\rm err}},
    \label{eq:trace_deviation_scaling}
\end{equation}
with $c\approx 0.1$ for the cases considered here. The approximately linear dependence of $\epsilon$ on $N$ differs from the $\sqrt{N}$ dependence of the first term in \cref{err_bound_normalize_formal}. The standard truncation estimate combines the discarded weights from different bonds in quadrature, giving rise to the factor $\sqrt{\sum_k\delta_k}$. In contrast, the trace deviations generated at different bonds are scalar contributions that can accumulate along a common axis, leading to an approximately linear dependence on the number of truncated bonds.

Substituting the empirical scaling in \eqref{eq:trace_deviation_scaling} into \cref{err_bound_normalize_formal}, we obtain the empirical prediction
\begin{align} \label{trunc_error_tr}
    \|\mathcal T_{\mathcal R}(\rho)-\rho\|_2
    &\lesssim
    f(N,\delta_{\rm err})\,\|\rho\|_2,
    \\
    f(N,\delta_{\rm err})
    &=
        \frac{
     \sqrt{\delta_{\rm err}} (\sqrt{2(N-1)}
    +cN)
    }{
    1-cN\sqrt{\delta_{\rm err}}
    }.
    \nonumber
\end{align}
Here we assume $cN\sqrt{\delta_{\rm err}}<1$.

In \cref{normalize_effect_fig}(c), we compute the observed error of the combined truncation-and-normalization operation and find that its scaling is in good agreement with \cref{trunc_error_tr}. In the remainder of this paper, we therefore use \cref{trunc_error_tr} as an empirical estimate of the single-step MPO compression error. We keep the coefficient $c$ explicit, as it can be calibrated for a given setup using the same numerical procedure illustrated in \cref{normalize_effect_fig}. For all numerical results presented in this paper, we set $c=0.1$.

\begin{figure}[h!]
	\centering
	\includegraphics[width=0.48\textwidth]{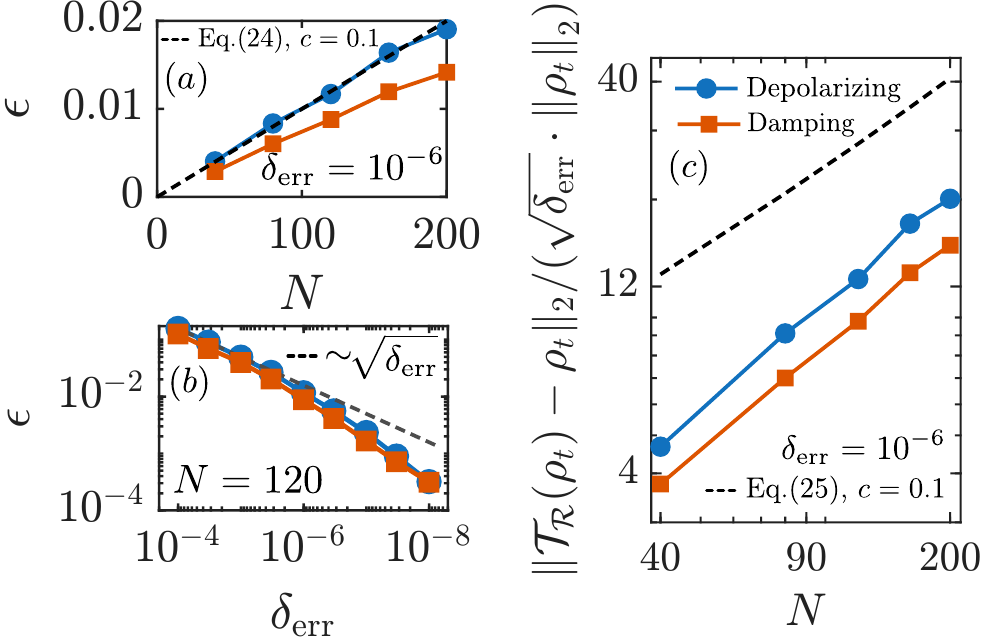}
\caption{
Single-step trace deviations $\epsilon$ and truncation-and-normalization errors for single-layer ($t=1$) 1D noisy random-circuit evolution under depolarizing noise ($p_{\rm dep}=0.02$, blue markers) and amplitude-damping noise ($p_{\rm damp}=0.1$, red markers) [see the setup in~\cref{circ_set}]. (a) Trace deviation $\epsilon$ as a function of system size $N$ at $\delta_{\rm err}=10^{-6}$, and (b) as a function of the truncation threshold $\delta_{\rm err}$ at $N=120$. (c) Relative $L_2$ error of the combined truncation-and-normalization operation as a function of $N$. The dashed lines in all panels indicate the corresponding empirical estimates.
}
\label{normalize_effect_fig}
\end{figure}

\subsubsection{Full truncation scheme}

For both the noisy random circuit [cf.~\cref{noi_evo}] and the Lindbladian dynamics [cf.~\cref{lind_evo}], truncation is applied after each layer of the physical evolution. Consequently, the overall $L_2$ error $\|\rho_t-\sigma_t\|_2$ accumulates and evolves with time $t$. We also point out that, although the above truncation scheme provides an empirical estimate of the $L_2$ error for each truncation [cf.~\cref{trunc_error_tr}], it does not directly ensure that the MPO simulation can be performed efficiently, as the bond dimension $D_k$ required to satisfy $\delta_k(D_k)\le \delta_{\rm err}$ may be very large. To ensure that the bond dimensions $\{D_k\}$ remain reasonably small during the evolution, a sufficient condition is a uniform area law for a R\'enyi-$\alpha$ operator entropy with $0<\alpha<1$ along the states being compressed; a von Neumann operator-entropy area law alone does not provide the required polynomial control of Schmidt tails~\cite{Verstraete2006c}. It has been shown that 1D random circuit dynamics with depolarizing noise or amplitude-damping noise exhibit area-law scaling of the operator entanglement~\cite{Noh2020,li2023entanglement,lee2025classical}, and similar area-law behavior has also been observed in 1D Lindbladian dynamics~\cite{PhysRevA.108.012616,PhysRevLett.129.170401}. In Appendix \ref{apd_area_sop}, we provide numerical evidence that the Lindbladian dynamics considered in this work also exhibit area-law scaling of the operator entanglement.

\subsubsection{Existing upper bound on the $L_1$ truncation error}
\label{l1_naive_bound}
Compared to the $L_2$ error $\|\rho_t - \sigma_t\|_2$, a physically more relevant measure is the $L_1$ error $\|\rho_t - \sigma_t\|_1$, as half this norm equals the trace distance when both operators are density matrices~\cite{Nielsen2000}, and bounds the TVD of their measurement distributions~\cite{hangleiter2023computational}. 
For a nonpositive MPO approximation, the $L_1$
error still controls the difference in measurement
weights, while obtaining a valid sampling distribution
requires an additional step.

However, for MPO-based simulation algorithms, it is typically difficult to obtain a meaningful bound on the $L_1$ error, since the difference $\rho_t-\sigma_t$ is often almost full rank, i.e., $\operatorname{rank}(\rho_t-\sigma_t)\sim 2^N$. Thus, \cref{L12_relation} leads to the following bound on the $L_1$ error made by the algorithm:
\begin{equation}\label{L1_l2_rat}
\|\rho_t-\sigma_t\|_1\leq \sqrt{2^N}\|\rho_t-\sigma_t\|_2.
\end{equation}
The exponentially large prefactor $\sqrt{2^N}$ renders this bound very loose for many-body density matrices. Therefore, even if one can bound $\|\rho_t-\sigma_t\|_2$ in a classical simulation algorithm using area-law scaling of operator entanglement~\cite{Noh2020,li2023entanglement,lee2025classical,PhysRevA.108.012616,PhysRevLett.129.170401,wei2025measurement,PhysRevA.104.022407,PhysRevA.108.052604}, such a bound need not yield a small $L_1$ error.

\begin{figure*}
	\centering
	\includegraphics[width=0.98\textwidth]{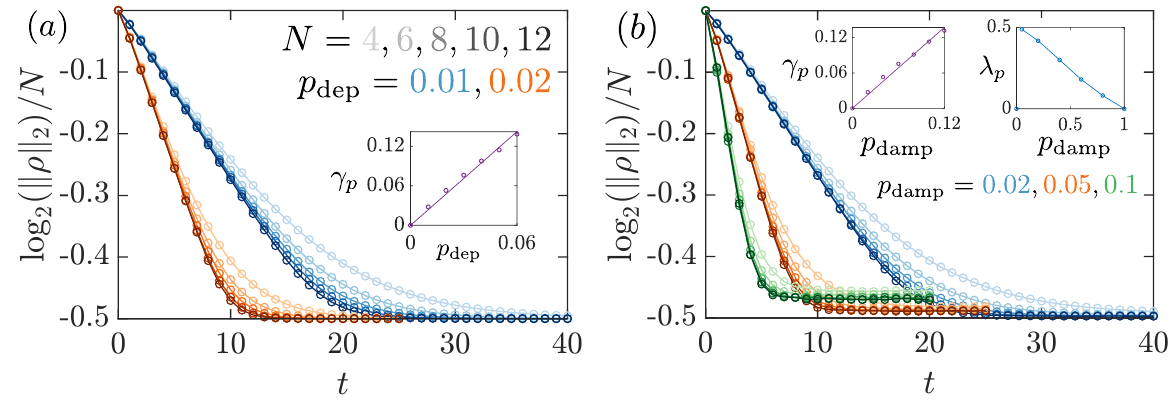}
        \caption{Evolution of the $L_2$ norm in noisy circuits, with system size $N=4,6,8,10,12$ (from lighter to darker colors). The number of circuit layers $t$ takes integer values. (a) Noisy circuits with depolarizing noise for various system sizes $N$ and noise rates $p_{\rm dep}$. The inset shows the extracted early-stage decay rate $\gamma_p$ [cf.~\cref{L2_scale_eq}] as a function of $p_{\rm dep}$. The fitted line is $\gamma_p = c_{\rm dep} p_{\rm dep}$ with $c_{\rm dep} \approx 2.37$.  (b) Same as panel (a) for amplitude-damping noise with rate $p_{\rm damp}$. In addition to the inset showing $\gamma_p$ and its fitted line $\gamma_p = c_{\rm damp} p_{\rm damp}$ with $c_{\rm damp}\approx 1.14$, we also plot, in a separate inset, $\lambda_p$, which characterizes the steady-state purity [cf.~\cref{L2_scale_eq}] as a function of $p_{\rm damp}$.}
        \label{l2nm_fig}
\end{figure*}

\subsection{Intuition on noise-induced error suppression}
\label{sec_pre_intu}

Taking noisy random circuit dynamics with depolarizing noise [cf.~\cref{circ_set}] as an example, we note that:
\begin{enumerate}
    \item The $L_2$ norm of the system density matrix $\|\rho_t\|_2$ [cf.~\cref{noi_evo}] decays exponentially with system size $N$ and time $t$ (\cite{aharonov1996limitations,deshpande2022tight,Boixo2018} and \cref{sec_norm}), before reaching the steady state. Thus the $L_2$ error introduced by each truncation can potentially be \textit{exponentially small} in both $N$ and $t$ [cf.~\cref{trunc_error_tr}].
    
    \item The distance between $\rho_t$ and $\sigma_t$ also evolves under the composite channel $\cal K$ [\cref{noi_evo,lind_evo}] as
$\rho_{t} - \sigma_{t} \mapsto {\cal K}(\rho_t - \sigma_t).$
Since the noisy dynamics can suppress differences between evolving operators, we expect it to reduce previously generated MPO truncation errors. This contraction mechanism does not require the state itself to converge to a fixed point; in particular, individual amplitude-damping random-circuit realizations remain time dependent at late times.
\end{enumerate}

In the remainder of this paper, we provide numerical evidence that this intuition holds for both 1D noisy random circuits [\cref{circ_num_sec}] and Lindbladian dynamics [\cref{lind_num_sec}], under both depolarizing (unital) and amplitude-damping (non-unital) noise, leading to empirical error estimates that are substantially tighter than the generic bound in \cref{L1_l2_rat}.

\section{Results for noisy random circuits}
\label{circ_num_sec}

In this section, we present the numerical and analytical results for the noisy random circuit setting [cf.~\cref{circ_set}].

\subsection{Behavior of the $L_2$ norm $\|\rho_t\|_2$}
\label{sec_norm}

Since the $L_2$ norm of the system density matrix $\|\rho_t\|_2=\sqrt{{\rm Tr}[\rho_t^2]}$  [cf.~\cref{norm_sec}] is closely related to the $L_2$ error introduced by MPO truncation [cf.~\cref{trunc_error_tr}], we begin by characterizing the evolution of $\|\rho_t\|_2$ under noisy circuit dynamics. To this end, we simulate the exact noisy circuit evolution and take an ensemble average over random circuits to obtain the average behavior of $\| \rho_t \|_2$. The system behavior converges after $\sim 100$ realizations. The same ensemble-averaging procedure is applied to all numerically computed quantities reported in \cref{circ_num_sec}, unless otherwise specified. 

Figures~\ref{l2nm_fig}(a,b) show the behavior of $\| \rho_t \|_2$ for the depolarizing and amplitude-damping cases, respectively, for $N=4,6,8,10,12$. We generally observe that the ensemble-averaged \(\|\rho_t\|_2\) relaxes toward a stationary late-time value. As $N$ increases, finite-size effects become less prominent, and the data approximately collapse toward the following behavior: 
\begin{equation} \label{L2_scale_eq}
\frac{1}{N}\log_2(\|\rho_t\|_2) \approx
\left\{
\begin{array}{ll}
-\gamma_p t, &  t\lesssim T_s, \\
-\lambda_p+O(e^{-s_{N,p} t}), &  t\gtrsim T_s.
\end{array}
\right.
\end{equation}

The $L_2$ norm thus decays exponentially with system
size $N$, with an initial temporal decay that crosses
over to a stationary late-time value. Here $\gamma_p$
characterizes the early-time decay, $\lambda_p$ the
stationary norm, and $s_{N,p}$ the late-time relaxation
rate of the logarithmic norm. The crossover timescale
is estimated as $T_s\simeq\lambda_p/\gamma_p$.

Note that the late-time correction in
\cref{L2_scale_eq} applies to the logarithmic norm.
Consequently, the relative deviation of
$\|\rho_t\|_2$ from its stationary value scales
as $O(Ne^{-s_{N,p}t})$ when this quantity is small.
We characterize $s_{N,p}$ in \cref{depo_sp_apd}
and neglect this small late-time correction in the
empirical error estimates.

In the case of depolarizing noise, the steady state is known to be the maximally mixed state for any \(p_{\rm dep} > 0\), which corresponds to \(\lambda_p = 1/2\). To determine \(\gamma_p\) for both noise models and \(\lambda_p\) for amplitude-damping noise, we note that the data collapse in \cref{l2nm_fig} indicates that the \(N=12\) results already provide a good proxy for the large-\(N\) behavior. We therefore numerically extract \(\gamma_p\) and \(\lambda_p \in [0, 1/2]\) as functions of the noise rates \(p_{\rm dep}\) and \(p_{\rm damp}\) from the \(N=12\) data. The resulting estimates are shown in the insets of \cref{l2nm_fig}. 

For both depolarizing noise and amplitude-damping noise, we observe an approximately linear relationship between the slope $\gamma_p$ and the corresponding noise rate $p_{\rm dep}$ or $p_{\rm damp}$, namely,
\begin{equation} \label{gamma_p_relation}
\begin{aligned}
&\gamma_p \approx c_{\rm dep}\, p_{\rm dep}, \qquad	c_{\rm dep} \approx 2.37
\\
&\gamma_p \approx c_{\rm damp} p_{\rm damp}, \quad c_{\rm damp}\approx 1.14.
\end{aligned}
\end{equation}
Note that the steady-state purity (characterized by $\lambda_p$) exhibits a qualitatively different behavior for the amplitude-damping case: for $p_{\rm damp}>0$, the steady-state purity increases monotonically with $p_{\rm damp}$, corresponding to $\lambda_p$ decreasing monotonically from $1/2$ toward $0$. This behavior arises from the interplay between random
unitary scrambling and the non-unital action of amplitude
damping; see \cref{amp_damp_mech,haar_purity_sec} for a
physical interpretation and a solvable many-body model.

\subsection{Noise-induced suppression of $L_2$ truncation error}
\label{sec_l2_err}
From \cref{L2_scale_eq}, the density matrix generated during the noisy circuit evolution up to arbitrary times \(t \ge 1\) has an \(L_2\) norm that decays exponentially with the system size \(N\). This implies that the empirical estimate of the single-step \(L_2\) truncation error is also exponentially small in \(N\) [cf.~\cref{trunc_error_tr}]. Moreover, as mentioned in \cref{sec_pre_intu}, we expect that the noisy circuit dynamics will further evolve the accumulated \(L_2\) truncation error. It is therefore an interesting question how the \(L_2\) error in the MPO simulation algorithm behaves under the combined action of these two effects, which is the subject of this section.

\subsubsection{Evolution of the single-step error}
The same noisy channel sequence acts on both the exact density matrix \(\rho_t\) and its truncated approximation \(\sigma_t\), and can contract the distance between them. We refer to this effect as noise-induced suppression of errors, and demonstrate it by studying the evolution of the $L_2$ error $\|\rho_t - \sigma_t\|_2$ throughout the noisy circuit evolution.

\textit{Noise-induced suppression for single-step $L_2$ error.} Here we study the evolution of single-step truncation errors, and show that the noisy random circuit evolution further suppresses the error.
We choose $\delta_{\rm err} = 10^{-6}$ in our study, and we have numerically verified that different $\delta_{\rm err}  \ll 1$ show qualitatively the same behavior.

\begin{figure}[t!]
	\centering
	\includegraphics[width=0.48\textwidth]{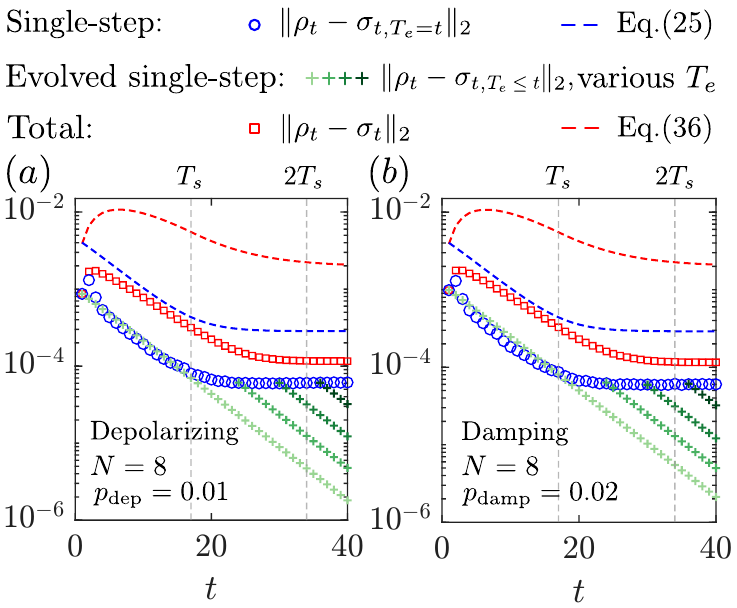}
        \caption{Behavior of $L_2$ truncation errors for noisy random circuits, for (a) depolarizing noise and (b) amplitude-damping noise. The number of circuit layers $t$ takes integer values. The blue circles denote the single-step error $\| \rho_t - \sigma_{t,T_e=t}\|_2$ [cf. \cref{L2_step_evo}], and the blue dashed line denotes the empirical single-step estimate of the truncation error \cref{trunc_error_tr}. The green markers (with multiple contrast levels) denote the evolution of the single-step error $\| \rho_t - \sigma_{t,T_e\leq t}\|_2$ for several single-time truncation locations $T_e=1,24,30,36$. The red squares denote the total truncation error $\| \rho_t - \sigma_{t}\|_2$, and the red dashed line denotes the empirical prediction in \cref{l2_scale_allT}. The vertical gray dashed lines indicate the number of layers $T_s$ (where the system almost reaches the steady state) and $2T_s$.
        }
        \label{L2err_fig}
\end{figure}

To probe the single-step truncation error at different stages of the noisy evolution, we consider the following setup in which the system density matrix is truncated only once, at a specific time $T_e$. This procedure produces an MPO denoted as $\sigma_{t,T_e}$, which evolves according to
\begin{align} \label{L2_step_evo}
\sigma_{t,T_e}=&\rho_t,\qquad \qquad \qquad \quad t<T_e \nonumber \\
\sigma_{t,T_e}=&({\cal T_R}\circ{\cal K}^{\cal C}_{t})(\sigma_{t-1,T_e}),\qquad t = T_e,\nonumber \\
\sigma_{t,T_e}=&{\cal K}^{\cal C}_{t}(\sigma_{t-1,T_e}), \qquad \qquad t > T_e.
\end{align}
The results are shown in \cref{L2err_fig} for both depolarizing noise and amplitude-damping noise. For fixed $N = 8$, we plot the empirical prediction in \cref{trunc_error_tr} as the dashed blue curve, which is proportional to $\| \rho_t \|_2$ and exhibits the same scaling behavior [cf.~\cref{L2_scale_eq}]. The blue circles in \cref{L2err_fig} show the behavior of $\| \rho_t - \sigma_{t,T_e=t} \|_2$ for different values of $t$. We observe that the single-step truncation error scales in close agreement with the prediction of the empirical estimate, aside from transient deviations at small $t$.
Together with the observed scaling with $N$, these results indicate that the empirical estimate in \cref{trunc_error_tr} correctly captures the scaling behavior of the truncation error.

The green markers denote $\|\rho_t - \sigma_{t,T_e\leq t}\|_2$ as a function of $t$ for several different values of $T_e$, illustrating the evolution of the error under noisy dynamics. We numerically observe that the truncation error is suppressed exponentially in $N$ at each evolution step, as for $t>T_e$,
\begin{equation} 
\begin{aligned}\label{err_shrink_pre}
\left\|\rho_t-\sigma_{t,T_e}\right\|_2
&=\left\|{\cal K}^{\cal C}_t\left(\rho_{t-1}-\sigma_{t-1,T_e}\right)\right\|_2 \\
&\approx 2^{-\gamma_p N}\left\|\rho_{t-1}-\sigma_{t-1,T_e}\right\|_2,
\end{aligned}
\end{equation}
where $\gamma_p$ is the same rate extracted from the decay of the $L_2$ norm [cf.~\cref{L2_scale_eq}].

\begin{figure}[t!]

	\centering
	\includegraphics[width=0.48\textwidth]{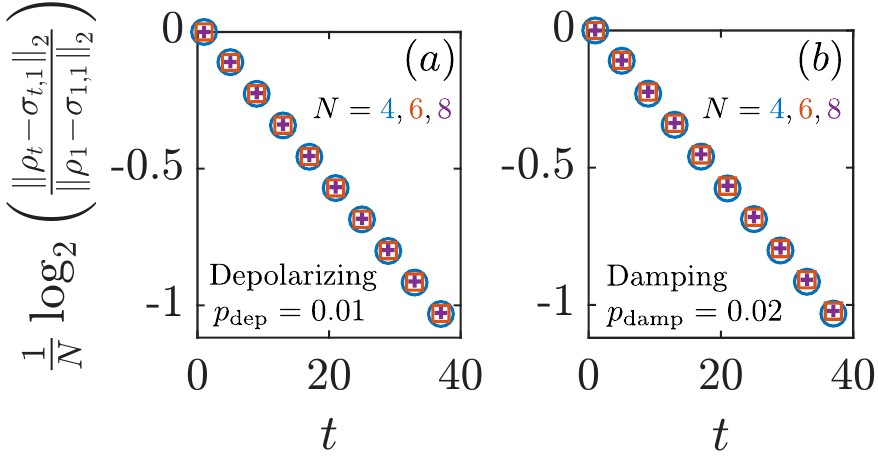}
        \caption{System-size scaling of the evolved single-step $L_2$ truncation error
under (a) depolarizing noise with $p_{\rm dep}=0.01$ and
(b) amplitude-damping noise with $p_{\rm damp}=0.02$.
A single truncation is introduced at $T_e=1$, and the quantity defined
in \cref{L2_shrink_calc_val} is plotted for $N=4,6,8$, represented by hollow circles,
hollow diamonds, and plus markers, respectively.}
        \label{L2_shrink_Nscale_fig}
\end{figure}

We independently verify the system-size scaling predicted by \cref{err_shrink_pre} by evaluating \begin{equation}  \label{L2_shrink_calc_val}
\frac{1}{N}\log_2\left( \frac{\left\|\rho_t-\sigma_{t,1}\right\|_2} {\left\|\rho_1-\sigma_{1,1}\right\|_2} \right). 
\end{equation} According to \cref{err_shrink_pre}, this quantity should be independent of $N$
and decrease linearly with 
$t$. Both predictions are clearly confirmed by the numerical results shown in \cref{L2_shrink_Nscale_fig}.

Based on \cref{err_shrink_pre}, we expect that the composite channel of the circuit and noise, ${\cal K}_t$, suppresses the $L_2$ error introduced by the MPO simulation algorithm in the same manner as it suppresses the $L_2$ norm of the system density matrix, namely,
\begin{equation} \label{err_shrink}
\left\|{\cal K}^{\cal C}_{t+1}\left(\rho_{t}-\sigma_{t}\right)\right\|_2 
\approx 2^{-\gamma_p N}\left\|\rho_{t}-\sigma_{t}\right\|_2.
\end{equation}
This observation is consistent with the intuition that noise reduces the distance between density matrices, thereby suppressing the truncation error in this setting.

The exponential dependence on $N$ in \cref{err_shrink_pre,err_shrink} is not a generic consequence of quantum-channel contractivity, since the $L_2$ norm need not decrease under an arbitrary channel.  The global random-unitary model in \cref{haar_toy_sec} makes this scaling analytically tractable. Within that model, the same channel-dependent factor governs both
the excess mean purity and the mean-squared norm of a traceless
error, providing an analytic counterpart to the common rate
$\gamma_p$ observed here.

\subsubsection{Evolution of the total error $\|\rho_t - \sigma_t\|_2$}
After understanding the behavior of the evolution of the single-step $L_2$ error, we are now ready to study the behavior of the total $L_2$ error $\|\rho_t - \sigma_t\|_2$. We focus on the regime of small truncation error $\delta_{\rm err}\ll 1$, which is the relevant regime for MPO simulation. In the following, we first construct an empirical scaling estimate for the total
error based on the previously observed scalings \cref{L2_scale_eq,trunc_error_tr,err_shrink}, and then numerically show that it captures the qualitative behavior
of the actual error.

\textit{Approximate analytical estimate.---\label{app_L2_predict}}To estimate $\left\Vert \rho_{t}-\sigma_{t}\right\Vert_{2}$ after $t$ layers of noisy circuit evolution, we apply the triangle inequality as
\begin{equation}
\label{tri_ineq}
\left\Vert \rho_{t}-\sigma_{t}\right\Vert _{2}\le
\left\Vert \rho_{t}-{\cal K}^{\cal C}_{t}(\sigma_{t-1})\right\Vert _{2}+
\left\Vert {\cal K}^{\cal C}_{t}(\sigma_{t-1})-\sigma_{t}\right\Vert _{2}.
\end{equation}
For the first term, we can utilize the noise-induced error suppression in \cref{err_shrink} to obtain
\begin{equation} 
	\left\Vert \rho_{t}-{\cal K}^{\cal C}_{t}(\sigma_{t-1})\right\Vert _{2} \approx 2^{-\gamma_p N} \left\Vert \rho_{t-1}-\sigma_{t-1}\right\Vert _{2},
\end{equation}
which transforms the term backward in time from $t$ to $t-1$. 
Here we use the contraction observed for
single-step truncation errors as an empirical
approximation for the propagation of accumulated
errors in the full MPO simulation. By repeatedly applying the same procedure (i.e., the triangle inequality and \cref{err_shrink}), we can express the total error as
\begin{equation} \label{L2scale_final}
\begin{aligned}
\left\Vert \rho_{t}-\sigma_{t}\right\Vert _{2}
&\lesssim \sum_{j=1}^{t} 2^{-\gamma_pN(t-j)}
    \left\Vert {\cal K}^{\cal C}_{j}(\sigma_{j-1})
      -{\cal T_R}\circ{\cal K}^{\cal C}_{j}(\sigma_{j-1}) \right\Vert _{2}\\
      &\lesssim f(N,\delta_{\rm err})
    \sum_{j=1}^{t} 2^{-\gamma_pN(t-j)}
    \left\Vert {\cal K}^{\cal C}_{j}(\sigma_{j-1}) \right\Vert _{2},
\end{aligned}
\end{equation}
where we used \cref{trunc_error_tr} in the second line. Moreover, when the single-step compression error
and the accumulated relative $L_2$ error remain
small, we have
$\|\mathcal K_j^{\cal C}(\sigma_{j-1})\|_2
\approx\|\sigma_j\|_2
\approx\|\rho_j\|_2$. Using the behavior of $\left\Vert \rho_{j} \right\Vert_{2}$ and omitting the exponentially small late-time correction [cf.~\cref{L2_scale_eq}], we obtain the empirical prediction for $\left\Vert \rho_{t}-\sigma_{t}\right\Vert _{2}$ as
\begin{widetext}
\begin{equation}\label{l2_scale_allT}
\left\Vert \rho_{t}-\sigma_{t}\right\Vert _{2} \lesssim 
f(N,\delta_{\rm err}) \,
\begin{cases}
t\cdot \,\left\Vert \rho_{t}\right\Vert _{2}, & t \lesssim T_s,\\[4pt]
\!\left(
2^{-\gamma_p N (t-T_s)}\,T_s
+\dfrac{1-2^{-\gamma_p N (t-T_s)}}{1-2^{-\gamma_p N}}
\right)\cdot \left\Vert \rho_{t}\right\Vert _{2}, & t\gtrsim T_s.
\end{cases}
\end{equation}
\end{widetext}

Therefore, in the regime $t\lesssim T_s$, the empirical estimate
predicts that the overall $L_2$ error inherits the exponential
suppression of the state norm,
$\|\rho_t\|_2\approx2^{-\gamma_pNt}$, multiplied by the additional
factor $t\,f(N,\delta_{\rm err})$. For $t\gtrsim T_s$, the system almost enters a steady state with $\left\Vert \rho_{t\gtrsim T_s}\right\Vert _{2}\approx 2^{-\lambda_pN}$. In the second line of \cref{l2_scale_allT}, the first term in parentheses shows that the noisy circuit evolution exponentially suppresses the errors accumulated during the $t\lesssim T_s$ regime, while the second term in parentheses captures the newly generated error for $t\gtrsim T_s$, which is likewise suppressed by the subsequent noisy evolution.  As a result, the empirical estimate predicts that the total truncation error approaches a saturated value
on a timescale $t\approx 2T_s$, rather than
accumulating indefinitely with circuit depth.

\textit{Numerical demonstration.} For both depolarizing and amplitude-damping noise, in \cref{L2err_fig} we plot the empirical prediction [cf.~\cref{l2_scale_allT}] as red dashed lines (where we use $T_s = \lfloor \lambda_p / \gamma_p \rfloor$ to determine which branch of the expression to plot), and show the actual total error $\left\Vert \rho_{t}-\sigma_{t}\right\Vert _{2}$ computed using the MPO simulation algorithm as red squares. Remarkably, the actual error exhibits the same qualitative behavior as the prediction. During the initial evolution, we observe an increase in the error, mainly due to the linearly increasing number of truncations. Subsequently, the error begins to decrease with $t$, indicating that noise-induced suppression starts to dominate the error evolution. After the timescale $T_s$, the system reaches its steady state, and the error is further reduced under the noisy circuit evolution, eventually reaching a steady value at $t\approx 2T_s$. We also observe that the actual error remains quantitatively below the empirical estimate in the cases studied, indicating that the accuracy of the MPO simulation algorithm is even better than expected from \cref{l2_scale_allT}. The quantitative discrepancy reflects both the conservativeness of the single-step empirical estimate in \cref{trunc_error_tr} for the cases studied and the triangle inequality in \cref{tri_ineq}.

\subsection{Evolution of $L_1$ error, and a potentially efficient MPO sampling algorithm}
\label{sec_l1_err}

\begin{figure*}[t!]
	\centering
	\includegraphics[width=\textwidth]{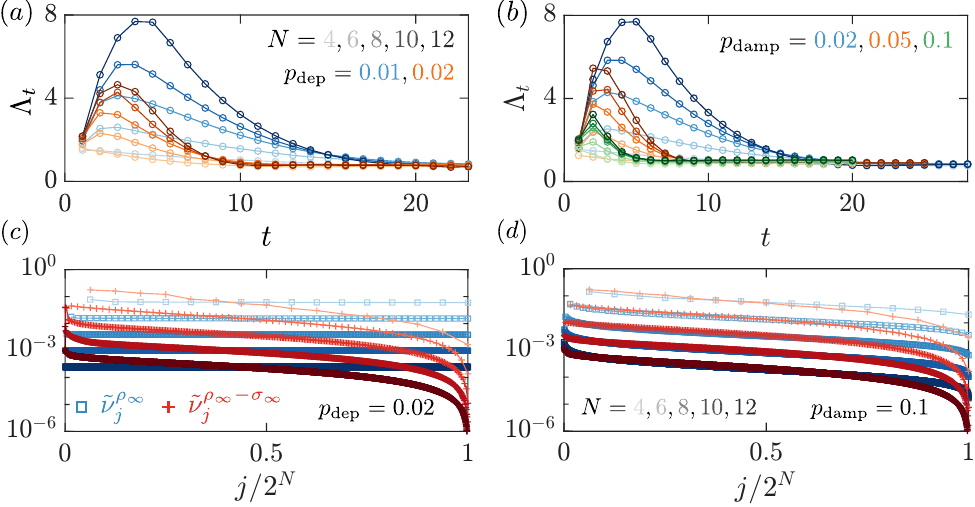}
        \caption{(a,b) Scaling behavior of the factor $\Lambda_t$ [\cref{lambda_t_defi}] for noisy random circuits, which controls the $L_1$ error $\|\rho_t-\sigma_t\|_1$ via \cref{l1_scale_allT}, for the case of (a) depolarizing noise and (b) amplitude-damping noise. The number of circuit layers $t$ takes integer values.  (c,d) Normalized singular-value spectrum ${\tilde{\nu}_j^{\rho_\infty}}$ and ${\tilde{\nu}_j^{\rho_\infty-\sigma_\infty}}$ of the steady state $\rho_\infty$ and of the corresponding steady-state error matrix $\rho_\infty-\sigma_\infty$, for (c) depolarizing noise and (d) amplitude-damping noise. In all panels, the system sizes are $N=4,6,8,10,12$, from lighter to darker colors.
        }
        \label{L1_rat_fig}
\end{figure*}

\subsubsection{Scaling behavior of $L_1$ error}

After understanding the behavior of $L_2$ errors, we are now equipped to study the behavior of the $L_1$ error. As shown in \cref{l2_scale_allT}, the $L_2$ error overall scales with the $L_2$ norm of the density matrix. For $\rho_t\ne\sigma_t$, we define
\begin{equation} \label{lambda_t_defi}
	\Lambda_t = \frac{\left\Vert \rho_{t}-\sigma_{t}\right\Vert _{1}}{\left\Vert \rho_{t}-\sigma_{t}\right\Vert _{2}}\cdot\left\Vert \rho_{t}\right\Vert _{2},
\end{equation}
which converts the relative $L_2$ error into the $L_1$ error. Substituting it into \cref{l2_scale_allT} gives
\begin{widetext}
\begin{equation}\label{l1_scale_allT}
\left\Vert \rho_{t}-\sigma_{t}\right\Vert _{1} \lesssim 
\Lambda_t \cdot f(N,\delta_{\rm err}) \,
\begin{cases}
t, & t \lesssim T_s,\\[4pt]
\!\left(
2^{-\gamma_p N (t-T_s)}\,T_s
+\dfrac{1-2^{-\gamma_p N (t-T_s)}}{1-2^{-\gamma_p N}}
\right), & t\gtrsim T_s.
\end{cases}
\end{equation}
\end{widetext}

At fixed nonzero noise strength and in the
small-truncation regime
$cN\sqrt{\delta_{\rm err}}\ll1$,
the remaining factors in \cref{l1_scale_allT}
are controlled by the truncation threshold
and the noise relaxation scale. Consequently, the scaling behavior of $\Lambda_t$ ultimately determines whether the MPO simulation algorithm can achieve an $L_1$ error that grows only polynomially with the system size $N$ (in the small-$L_1$-error regime).

We numerically investigate the scaling of $\Lambda_t$ in \cref{L1_rat_fig}(a,b) for system sizes $N=4,6,8,10,12$, considering both depolarizing and amplitude-damping noise. 
We observe a common behavior across different noise types and noise rates: for numbers of circuit layers $t \lesssim T_s$, $\Lambda_t$ initially grows and then decreases, and its value at a fixed $t$ increases with the system size $N$. Since this regime corresponds to the interplay between quantum correlations and noise, the observed behavior indicates that the $L_1$ simulation error exhibits a nontrivial system-size dependence beyond the explicitly retained factor $f(N,\delta_{\rm err})$ in \cref{l1_scale_allT}. 
More interestingly, for $t\gtrsim T_s$, where the
system approaches its stationary late-time regime,
we find that $\Lambda_t$ appears independent of
system size over the accessible sizes,
\begin{equation}\label{L1_collapse}
\Lambda_{t\gtrsim T_s}\approx\Lambda_\infty,
\qquad
\Lambda_\infty=O(1)
\quad\text{(empirical scaling)}.
\end{equation}

It is useful to distinguish how this result compares
with the generic rank bound in the two noise models.
For depolarizing noise, $\|\rho_\infty\|_2\approx
2^{-N/2}$, so $\Lambda_\infty=O(1)$ implies
$\|\rho_\infty-\sigma_\infty\|_1/
\|\rho_\infty-\sigma_\infty\|_2=O(2^{N/2})$.
Thus, the late-time $L_1/L_2$ conversion does not
improve the exponential prefactor of the generic
rank bound in \cref{L1_l2_rat}; instead, the
improvement comes from the noise-induced suppression
of the underlying $L_2$ truncation error
[cf.~\cref{l2_scale_allT}].

By contrast, for amplitude-damping noise,
$\|\rho_\infty\|_2\sim2^{-\lambda_pN}$
with $\lambda_p<1/2$ [cf.~\cref{L2_scale_eq}],
giving
$\|\rho_\infty-\sigma_\infty\|_1/
\|\rho_\infty-\sigma_\infty\|_2
=O(2^{\lambda_pN})$,
which is exponentially smaller than the generic
$2^{N/2}$ prefactor.

\subsubsection{Spectral mechanism underlying the late-time \(L_1/L_2\) error conversion factor}
\label{spectrum_sec}
To clarify the observed late-time scaling
$\Lambda_\infty=O(1)$ [cf.~\cref{L1_collapse}],
we analyze the singular-value spectra of the
density matrix $\rho_\infty$ and the corresponding
error matrix $\rho_\infty-\sigma_\infty$.
Here, $\rho_\infty$ and $\sigma_\infty$ denote
representative states in the stationary late-time
regime with nonzero truncation error, rather than
strict infinite-time limits.

Consider a matrix $A$ whose singular values ${\nu_j^{A}}$ are arranged in descending order, $\nu_1^{A} \ge \nu_2^{A} \ge \cdots\ge 0$. The ratio between its $L_1$ and $L_2$ norms can then be written as
\begin{equation} 
\frac{\|A\|_1}{\|A\|_2}=\frac{\sum_j \nu_j^A}{\sqrt{\sum_j (\nu_j^A)^2}}
=\frac{1}{\sqrt{\sum_j (\tilde{\nu}_j^A)^2}},
\end{equation}
where the normalized singular values are defined by $\tilde{\nu}_j^A = \nu_j^A / \sum_j \nu_j^A$. Therefore, for an arbitrary number of circuit layers $t$, $\Lambda_t$ [cf.~\cref{lambda_t_defi}] is fully determined by the normalized singular-value spectrum of $\rho_t-\sigma_t$ and $\rho_t$, denoted by $\tilde{\nu}_j^{\rho_t-\sigma_t}$ and $\tilde{\nu}_j^{\rho_t}$, respectively, as
\begin{equation} \label{lambda_t_eq}
	\Lambda_t=\sqrt{\frac{\sum_j(\tilde\nu_j^{\rho_t})^2}{\sum_j(\tilde\nu_j^{\rho_t-\sigma_t})^2}}.
\end{equation}

Therefore, $\Lambda_\infty$ can be interpreted as a measure of the relative flatness of the normalized singular-value spectrum ${\tilde{\nu}_j^{\rho_\infty}}$ and ${\tilde{\nu}_j^{\rho_\infty-\sigma_\infty}}$ of the steady state $\rho_\infty$ and of the corresponding steady-state error matrix $\rho_\infty-\sigma_\infty$. In particular, $\Lambda_\infty < 1$ when the singular-value spectrum of $\rho_\infty$ is flatter than that of $\rho_\infty-\sigma_\infty$, whereas $\Lambda_\infty \approx 1$ when the two spectra have similar shapes.

In \cref{L1_rat_fig}(c,d), we show the representative behavior of the normalized steady-state singular-value spectrum of both $\rho_{\infty}$ and $\rho_{\infty}-\sigma_{\infty}$ for depolarizing noise with rate $p_{\rm dep}=0.02$ and amplitude-damping noise with rate $p_{\rm damp}=0.1$. 
Because the singular values at a fixed index $j$
can fluctuate by more than an order of magnitude
across random realizations, we use geometric
averaging to display a typical spectral profile.
The resulting averaged spectra are not necessarily
normalized and cannot be directly substituted
into \cref{lambda_t_eq}.

For the depolarizing case [cf.~\cref{L1_rat_fig}(c)], the late-time state $\rho_\infty$ is close to maximally mixed, and its spectrum is approximately flat. At the same time, the spectrum of the error matrix $\rho_\infty-\sigma_\infty$ is also broad, retaining appreciable weight over most indices $1 \le j \le 2^N$. Moreover, both spectra exhibit qualitatively similar shapes across different system sizes $N$.

By contrast, for the amplitude-damping case [cf.~\cref{L1_rat_fig}(d)], the steady state is no longer maximally mixed, and the spectrum of $\rho_\infty$ therefore displays a nontrivial decay profile. We likewise find that the spectrum of $\rho_\infty-\sigma_\infty$ exhibits a qualitatively similar decay profile to that of $\rho_\infty$. As in the depolarizing case, the shape of each spectrum remains qualitatively unchanged as the system size increases.

The above observation---that the normalized singular-value spectra
of both the steady state $\rho_\infty$ and the corresponding error
matrix $\rho_\infty-\sigma_\infty$ retain qualitatively similar
shapes across the accessible system sizes---is consistent with the
observed size-independent behavior of $\Lambda_\infty$.
Moreover, the similarity between these two spectra further suggests that $\Lambda_\infty \approx 1$, in agreement with the numerical results shown in \cref{L1_rat_fig}(a,b).

The similarity between the spectral shapes of the steady state and its corresponding error matrix can be qualitatively understood as follows.
(1) Owing to noise-induced error suppression [cf.~\cref{sec_l2_err}], the truncation error accumulated before the system reaches its steady state is damped away [cf.~\cref{l2_scale_allT}]. As a result, the steady-state error matrix $\rho_{\infty}-\sigma_{\infty}$ is generated predominantly by the single-step truncation acting on the steady state.
(2) Each single-step MPO truncation discards the components corresponding to the smaller operator-Schmidt values across the MPO bonds~\cite{Verstraete2006c,baumgratz2014efficient}. The resulting single-step error is therefore composed of discarded
components of the same MPO representation of the current density
matrix. This provides a qualitative mechanism by which the error
matrix can retain spectral features of $\rho_\infty$, consistent
with our numerical observations.

\subsubsection{Potentially efficient classical sampling from noisy random 1D circuits using MPO}
\label{mpo_circ_samp_algo}
The numerical and analytical results above motivate the empirical
$L_1$-error estimate in \cref{l1_scale_allT}. In addition, our
numerical results indicate that the factor $\Lambda_t$ becomes
approximately independent of system size once the system enters
its stationary late-time regime [cf.~\cref{L1_collapse}].
Under these empirically observed scalings, we now construct an MPO
simulation strategy that is expected to sample from states generated
by 1D noisy random-circuit evolution with small \emph{total variation
distance} (TVD) from the true distribution.

Since MPO truncation does not necessarily preserve positivity,
we first address how to sample from a potentially nonpositive
MPO approximation. Define
$p(x)=\langle x|\rho_t|x\rangle$ and
$q(x)=\langle x|\sigma_{t,H}|x\rangle$, where
$\sigma_{t,H}=(\sigma_t+\sigma_t^\dagger)/2$.
If $q(x)\geq0$ for all $x$, standard sequential MPO sampling
produces the distribution $q$, with
${\rm TVD}(p,q)\leq\|\rho_t-\sigma_t\|_1/2$.

For general signed weights, define the prefix sums
$q(y)=\sum_zq(yz)$ and sample each subsequent bit $b$
according to
\begin{equation}
Q(b|y)=\frac{[q(yb)]_+}{[q(y0)]_++[q(y1)]_+},
\qquad [u]_+=\max(u,0),
\label{je_signed_sampler}
\end{equation}
using any probability distribution if the denominator
vanishes. Each prefix sum can be evaluated by MPO
contraction. The resulting sampling distribution $Q$
satisfies
\begin{equation}
{\rm TVD}(p,Q)
\leq N\sum_x|p(x)-q(x)|
\leq N\|\rho_t-\sigma_t\|_1,
\label{je_signed_sampling_bound}
\end{equation}
as proved in Appendix~\ref{je_sampling_appendix}.
Thus, a small $L_1$ error provides a constructive
sampling guarantee even without positivity,
at the cost of an additional factor $N$.

We consider the following strategy to sample from the output of 1D noisy random circuits acting on $N$ qubits. For a circuit of $t = O(\log N)$ layers, one can run the MPO simulation algorithm [cf.~\cref{sec_MPO_algo}] without incurring any truncation error ($\delta_{\rm err}=0$), with a runtime that scales polynomially in $N$, with ${\rm TVD}_{t=O(\log N)}=0$.

At fixed nonzero noise strength, for circuit depths
$t=\Omega(\log N)$ in the stationary late-time regime,
combining \cref{l1_scale_allT,L1_collapse} gives,
for nonnegative MPO output weights,
\begin{equation} \label{circ_L1_tvd}
{\rm TVD}_{t=\Omega(\log N)}\lesssim f(N,\delta_{\rm err}) \left(
\,T_s
+\dfrac{1}{1-2^{-\gamma_p N}}
\right) \Lambda_{\infty}.	
\end{equation}
Consider the regime where $f(N,\delta_{\rm err})\ll 1$, which can be achieved by choosing a sufficiently small $\delta_{\rm err}$. In this case, we have $f(N,\delta_{\rm err})= O(N \sqrt{\delta_{\rm err}})$ [cf.~\cref{trunc_error_tr}]; 
hence, choosing
$\delta_{\rm err}=O(\varepsilon^2/N^2)$
with a sufficiently small prefactor yields
a target TVD $\varepsilon$ for nonnegative
output weights, under the empirical assumptions.
For the general signed-weight sampler in
\cref{je_signed_sampler}, the conservative
choice is instead
$\delta_{\rm err}=O(\varepsilon^2/N^4)$. Under the additional assumption of a uniform area law for the R\'enyi-$\alpha$ operator entropy, with $0<\alpha<1$, along the normalized vectorizations of all inputs actually compressed by the algorithm, consistent with but stronger than the available entanglement evidence~\cite{Noh2020,li2023entanglement,lee2025classical}, the maximal MPO bond dimension over the course of the evolution scales as
$D_{\rm max}=O\left(\delta_{\rm err}^{\frac{-\alpha}{1-\alpha}}\right)$
~\cite{Verstraete2006c}. Summarizing, we expect the MPO algorithm to sample from $\rho_t$ at arbitrary circuit layer number $t$ with a small constant total variation distance, with runtime polynomial in $N$ and the number of supplied circuit layers $t$, at fixed noise strength and under the stated uniform error and Schmidt-tail assumptions.

\textit{Potential Implications.} The above simulation strategy is particularly interesting in the case of amplitude-damping noise for circuit evolutions of $t=\Omega(\log N)$ layers. For this regime, it was previously unknown whether the MPO simulation algorithm is efficient. 
Recent work has also established efficient
estimation of local observables under non-unital
noise~\cite{mele2026noise}.
These results are complementary to ours, as
efficient local-observable estimation does not
by itself guarantee efficient sampling from
the full output distribution with small TVD.
 Moreover, although Pauli-path based simulation methods
\cite{10.1145/3564246.3585234,schuster2024polynomial} have rigorous guarantees
in several noisy-circuit regimes, such guarantees are not known to apply here,
due to the absence of anti-concentration in the output distribution under
non-unital noise \cite{PRXQuantum.5.030317}. A ``patching'' algorithm~\cite{napp2022,lee2025classical} is expected to be efficient in this regime~\cite{lee2025classical}; however, we are not aware of an existing numerical implementation of the ``patching'' algorithm, suggesting that it may be less practical than MPO-based simulation algorithms. Our results therefore suggest that the widely used MPO simulation approach may already provide a potentially efficient simulation strategy in this setting, once the noise-induced suppression of truncation errors is taken into account. More broadly, they indicate that the non-unital nature of the noise does not necessarily render tensor-network–based classical simulation qualitatively more difficult than in the unital-noise case~\cite{PRXQuantum.5.030317}.


\section{A solvable global random-unitary model for error contraction}
\label{haar_toy_sec}

The local random-circuit numerics in
\cref{sec_norm,sec_l2_err} revealed that two apparently different
quantities exhibit approximately the same transient decay exponent:
the $L_2$ norm of the state and the $L_2$ norm of a previously generated MPO
truncation error. We now show that the same coincidence occurs exactly in a solvable global random-unitary model.

The solvable model is designed to isolate error propagation from error
generation. We replace the local unitary layers of the brickwall
circuit by independent global random unitaries, while keeping the same
physical product-noise channel. No MPO truncation is performed within
the solvable model itself. When discussing truncation errors, the input is
instead a previously generated truncation error by the local MPO algorithm of
\cref{sec_l2_err}. Thus, this solvable model is not intended as a
quantitative model of the local circuit, but as a solvable
second-moment model for the mechanism underlying the observed error
contraction.

\subsection{Common Haar-averaged contraction factor}
\label{haar_mean_square_sec}

Consider $N$ qubits with Hilbert-space dimension $d=2^N$, and define
one scrambled-noise substep by
\begin{equation}
\mathcal S_U(A)
:=
{\cal N}\!\left(UAU^\dagger\right),
\quad
U\sim{\rm Haar}[U(d)],
\quad
{\cal N}=\phi^{\otimes N},
\label{haar_substep}
\end{equation}
where $\phi$ is the single-qubit noise channel. To mirror the two
noise applications in each full brickwall layer of \cref{circ_set},
we define one full global random-unitary layer as
\begin{equation}
\mathcal K_t^{\rm H}
:=
\mathcal S_{U_{t,2}}
\circ
\mathcal S_{U_{t,1}},
\qquad
U_{t,1},U_{t,2}
\stackrel{\rm iid}{\sim}
{\rm Haar}[U(d)].
\label{haar_full_layer}
\end{equation}

For any fixed traceless operator $X$, $\operatorname{Tr}X=0$, Haar
isotropy implies the standard second-moment identity~\cite{collins2006integration}
\begin{equation}
\mathbb E_U
\left\|
{\cal N}(UXU^\dagger)
\right\|_2^2
=
\chi_{\cal N}
\left\|X\right\|_2^2,
\label{haar_chi_def}
\end{equation}
where
\begin{equation}
\chi_{\cal N}
=
\frac{1}{d^2-1}
\sum_{a=1}^{d^2-1}
\left\|
{\cal N}(F_a)
\right\|_2^2.
\label{haar_channel_average}
\end{equation}

Equivalently, include $F_0=\mathbb I_d/\sqrt d$ in the orthonormal basis and write
\begin{equation}
\chi_{\mathcal N}=\frac{\|\mathcal N\|_{\rm F}^2-d b_{\mathcal N}}{d^2-1},
\qquad b_{\mathcal N}=\left\|\mathcal N\left(\frac{\mathbb I_d}{d}\right)\right\|_2^2,
\label{je_channel_frobenius}
\end{equation}
where $\|\mathcal N\|_{\rm F}^2=\sum_{a=0}^{d^2-1}\|\mathcal N(F_a)\|_2^2$ is the squared Frobenius norm of the superoperator. For product noise, both terms factorize into single-site quantities.
Here $\{F_a\}_{a=1}^{d^2-1}$ is any orthonormal basis of the
traceless operator subspace. Hence, $\chi_{\cal N}$ is the squared
Frobenius action of the noise channel averaged over traceless input
directions.

To see this, vectorize operators in Hilbert--Schmidt space. Haar averaging of $|UXU^\dagger\rangle\langle UXU^\dagger|$ gives $\|X\|_2^2\Pi_0/(d^2-1)$, where $\Pi_0$ projects onto traceless operators; applying the channel and taking the trace yields \cref{haar_chi_def}. Only the unitary second moment is used, so an independent exact unitary $2$-design gives the same identities.

Because ${\cal N}$ is trace preserving, it maps traceless operators
to traceless operators. Applying \cref{haar_chi_def} to the two
independent substeps in \cref{haar_full_layer} therefore gives
\begin{equation}
\mathbb E_{U_{t,1},U_{t,2}}
\!\left[
\left\|
\mathcal K_t^{\rm H}(X)
\right\|_2^2
\,\middle|\,
X
\right]
=
\chi_{\cal N}^{\,2}
\left\|X\right\|_2^2.
\label{haar_full_layer_traceless}
\end{equation}
Thus, $\chi_{\cal N}$ is the root-mean-square contraction factor per
full global random-unitary layer.

\subsection{The same factor governs transient purity and error propagation}
\label{haar_purity_sec}

We now apply \cref{haar_full_layer_traceless} to the two quantities
that exhibited a common decay rate in the local-circuit numerics.

First consider the globally scrambled state
$\varrho_{t+1}=\mathcal K_t^{\rm H}(\varrho_t)$, and define its
ensemble-averaged purity
\begin{equation}
P_t
:=
\mathbb E\,\operatorname{Tr}(\varrho_t^2).
\end{equation}
Writing
$\varrho=\mathbb I_d/d+X$ with $\operatorname{Tr}X=0$ and applying
the Haar second-moment identity shows that the purity evolves by an
affine map whose deviation from its stationary value $P_{\rm ss}$
is multiplied by $\chi_{\cal N}$ at each substep. Consequently,
the one-substep mean-purity update and its fixed point are explicitly
\begin{equation}
\begin{aligned}
P'&=b_{\mathcal N}+\chi_{\mathcal N}(P-1/d),\\
P_{\rm ss}&=\frac{b_{\mathcal N}-\chi_{\mathcal N}/d}{1-\chi_{\mathcal N}},
\qquad \chi_{\mathcal N}<1.
\end{aligned}
\label{je_exact_stationary_purity}
\end{equation}
The cross term vanishes because $\mathbb E_U[UXU^\dagger]=0$. Iterating this affine recurrence over full layers gives
\begin{equation}
P_t-P_{\rm ss}
=
\chi_{\cal N}^{\,2t}
\left(P_0-P_{\rm ss}\right).
\label{haar_purity_solution}
\end{equation}

Next, suppose that the local MPO algorithm generates a truncation
error at time $T_e$,
\begin{equation}
E_0
:=
\rho_{T_e}-\sigma_{T_e}.
\label{haar_error_initial}
\end{equation}
Since both $\rho_{T_e}$ and the normalized MPO approximation
$\sigma_{T_e}$ have unit trace,
$\operatorname{Tr}E_0=0$. The global random-unitary model begins after
this error has been generated and propagates it without further
truncation,
\begin{equation}
E_{t+1}
=
\mathcal K_t^{\rm H}(E_t).
\label{haar_error_evolution}
\end{equation}
Iterating \cref{haar_full_layer_traceless} over independent layers
then gives
\begin{equation}
\mathbb E\left\|E_t\right\|_2^2
=
\chi_{\cal N}^{\,2t}
\left\|E_0\right\|_2^2.
\label{haar_fresh_composition}
\end{equation}
The same statement holds conditionally when $E_0$ is itself random,
provided that the subsequent Haar unitaries are fresh and
independent of the history that generated $E_0$.

The central result of the solvable model can therefore be summarized as
\begin{equation}
\begin{aligned}
\sqrt{|P_t-P_{\rm ss}|}
&=
\chi_{\cal N}^{\,t}
\sqrt{|P_0-P_{\rm ss}|},
\\
\sqrt{
\mathbb E\left\|E_t\right\|_2^2
}
&=
\chi_{\cal N}^{\,t}
\left\|E_0\right\|_2 .
\end{aligned}
\label{haar_common_rms_decay}
\end{equation}
When the transient contribution dominates the stationary floor,
\(P_t-P_{\rm ss}\simeq P_t\), so
\(\sqrt{|P_t-P_{\rm ss}|}\simeq\sqrt{\mathbb E\|\varrho_t\|_2^2}\). Thus, in this
regime, the RMS state \(L_2\) norm and the RMS norm of a pre-existing
error decay with the same factor \(\chi_{\cal N}\). More generally,
within the global random-unitary model, the excess mean purity and the
mean-squared norm of any initial traceless error are governed
by the same channel-dependent factor \(\chi_{\cal N}^2\) per full
layer. This provides an analytic counterpart to the common decay
exponent observed in
\cref{L2_scale_eq,err_shrink_pre}. Importantly, the propagation
result depends on the initial error only through
\(\|E_0\|_2\), and not on the details of the MPO truncation that
generated it.

\subsection{Channel-specific predictions and comparison with local circuits}
\label{haar_channel_results_sec}

We now evaluate $\chi_{\cal N}$ for the two noise channels used in
the local-circuit simulations.

For depolarizing noise, with
\begin{equation}
q_{\rm dep}
:=
1-\frac{4p_{\rm dep}}{3},
\end{equation}
one obtains the exact finite-$N$ factor
\begin{equation}
\chi_{\rm dep}(N,p_{\rm dep})
=
\frac{
(1+3q_{\rm dep}^2)^N-1
}{
4^N-1
}.
\label{haar_dep_chi}
\end{equation}
For fixed $p_{\rm dep}>0$, excluding the
completely depolarizing point $p_{\rm dep}=3/4$,
the large-$N$ behavior is
\begin{equation} 
	\chi_{\rm dep}
=
2^{-\gamma_{\rm dep}^{\rm Haar}N+o(N)},
\end{equation}
with
\begin{equation}
\gamma_{\rm dep}^{\rm Haar}
=
-\log_2\!\left(
\frac{1+3q_{\rm dep}^2}{4}
\right)
=
\frac{2}{\ln2}p_{\rm dep}
+O(p_{\rm dep}^2).
\label{haar_dep_exponent}
\end{equation}
Since depolarizing noise is unital,
\begin{equation}
P_{\rm ss}^{\rm dep}
=
\frac{1}{d}.
\label{haar_dep_purity_ss}
\end{equation}

For amplitude-damping noise, direct evaluation gives
\begin{equation}
\begin{aligned}
\chi_{\rm damp}(N,p_{\rm damp})
&=
\frac{
4^N\bigl(1-p_{\rm damp}+\frac{1}{2}p_{\rm damp}^2\bigr)^N
}{
4^N-1
}
\\
&\quad
-
\frac{
\bigl(1+p_{\rm damp}^2\bigr)^N
}{
4^N-1
}.
\end{aligned}
\label{haar_damp_chi}
\end{equation}

For amplitude damping, using
$\|\phi\|_{\rm F}^2=4-4p+2p^2$
and
$\|\phi(\mathbb I/2)\|_2^2=(1+p^2)/2$,
where $p=p_{\rm damp}$, we obtain the exact
finite-$N$ stationary mean purity
\begin{equation}
P_{\rm ss}^{\rm damp}
=
\frac{[(1+p^2)/2]^N
-2^{-N}\chi_{\rm damp}}
{1-\chi_{\rm damp}},
\qquad 0<p\leq1.
\label{je_damping_purity_finiteN}
\end{equation}

For fixed $0<p_{\rm damp}<1$ and large $N$,
\begin{equation} 
\chi_{\rm damp}
=
2^{-\gamma_{\rm damp}^{\rm Haar}N+o(N)},
\end{equation}
with
\begin{equation}
\begin{aligned}
\gamma_{\rm damp}^{\rm Haar}
&=
-\log_2\!\left(
1-p_{\rm damp}
+\frac12p_{\rm damp}^2
\right) \\
&=
\frac{1}{\ln2}p_{\rm damp}
+O(p_{\rm damp}^3).
\end{aligned}
\label{haar_damp_exponent}
\end{equation}
Unlike the depolarizing case, amplitude damping is non-unital and
produces a nontrivial stationary purity. Its large-$N$ behavior is
\begin{equation} 
	\sqrt{P_{\rm ss}^{\rm damp}}
=
2^{-\lambda_{\rm damp}^{\rm Haar}N+o(N)},
\end{equation}
with
\begin{equation}
\lambda_{\rm damp}^{\rm Haar}(p_{\rm damp})
=
\frac12
\log_2\!\left(
\frac{2}{1+p_{\rm damp}^2}
\right).
\label{haar_damp_purity_largeN}
\end{equation}

For a pure initial state, the crossover between the transient and
stationary contributions occurs on an inverse-noise timescale. In
the weak-noise, large-$N$ limit,
\begin{equation}
T_s^{\rm dep}
\sim
\frac{\ln2}{4p_{\rm dep}},
\qquad
T_s^{\rm damp}
\sim
\frac{\ln2}{2p_{\rm damp}}.
\label{haar_crossover_scales}
\end{equation}

\begin{figure}[t!]
\centering
\includegraphics[width=0.48\textwidth]
{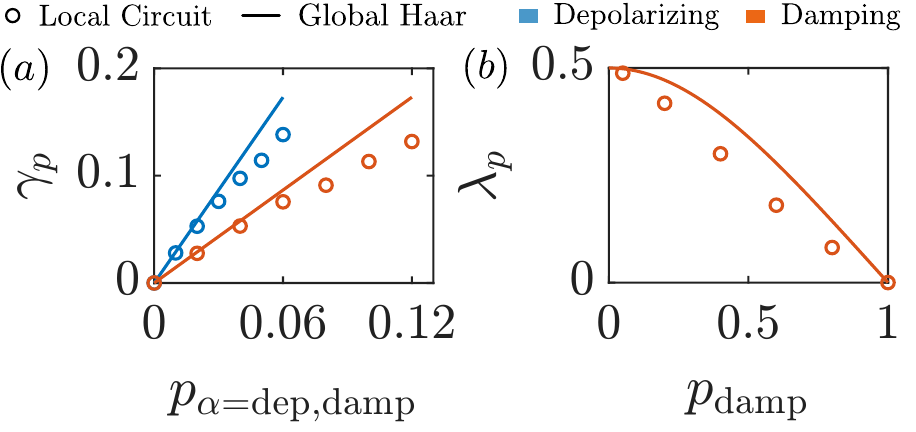}
\caption{
Comparison between the local brickwall circuit and the global random-unitary
solvable model. Open circles denote the results of local-circuit numerics, while 
solid
lines or curves denote global random-unitary predictions; the local-circuit
data are the same as those shown 
in the insets of
\cref{l2nm_fig}.
(a) Full-layer contraction exponent for depolarizing and
amplitude-damping noise. The straight lines show the leading
weak-noise predictions in
\cref{haar_dep_exponent,haar_damp_exponent}.
(b) Stationary $L_2$-norm exponent for amplitude-damping noise,
compared with $\lambda_{\rm damp}^{\rm Haar}$ in
\cref{haar_damp_purity_largeN}.
}
\label{local_vs_haar_fig}
\end{figure}

As shown in \cref{local_vs_haar_fig}(a), the global random-unitary weak-noise
predictions reproduce well the contraction rates extracted from the
local brickwall circuits at the smallest noise strengths studied.
For amplitude damping, \cref{local_vs_haar_fig}(b) further shows that
the solvable model captures the monotonic dependence and endpoint limits
of the stationary $L_2$-norm exponent, although quantitative
deviations remain at intermediate damping strengths.

Taken together, the solvable model accounts for the main qualitative
features of the local-circuit numerics: exponential-in-$N$
contraction of a pre-existing error, a common transient second-moment
exponent for state purity and error propagation, an inverse-noise
crossover scale, and a nontrivial stationary purity for non-unital
noise. The agreement should not be interpreted as an exact mapping:
global random scrambling removes the locality and finite-depth
structure of the brickwall circuit, and the solvable model does not
describe the adaptive generation of MPO truncation errors. Rather,
it isolates a simple channel-level mechanism by which the same
second-moment factor can control both state mixing and the
suppression of a pre-existing truncation error.


\section{Results for Trotterized Lindbladian Dynamics}
\label{lind_num_sec}

The global random-unitary analysis in \cref{haar_toy_sec} gives a solvable mechanism by which product noise can contract a globally scrambled traceless error exponentially in system size, and it relates this contraction to the transient state purity.  Lindbladian dynamics nevertheless differ in important ways from both the global random-unitary model and noisy random circuits because their coherent evolution is deterministic, local, and continuous.  In this section, we therefore test whether the same truncation-error phenomenology persists in an interacting non-integrable Lindbladian setting without global random scrambling.  We carry out an analysis analogous to the random-circuit case [cf.~\cref{circ_num_sec}] and present the discussion more briefly when describing analogous phenomena.

In the following, we use MPO to simulate the Trotterized Lindbladian evolution [\cref{lind_set}]. The Hamiltonian parameters and noise rates are $(g,h,\kappa_{\rm dep})=(1,1,0.05 \textrm{ or } 0.1)$ for depolarizing noise, and $(g,h,\kappa_{\rm damp})=(1 \textrm{ or } 8,1,0.4)$ for amplitude-damping noise. We use a discretization step of $\Delta t=0.05$ and an MPO truncation threshold of $\delta_{\rm err}=10^{-6}$ throughout the simulation.

\subsection{Behavior of the $L_2$ norm $\|\rho_t\|_2$}

\begin{figure}[t!]
	\centering
	\includegraphics[width=0.48\textwidth]{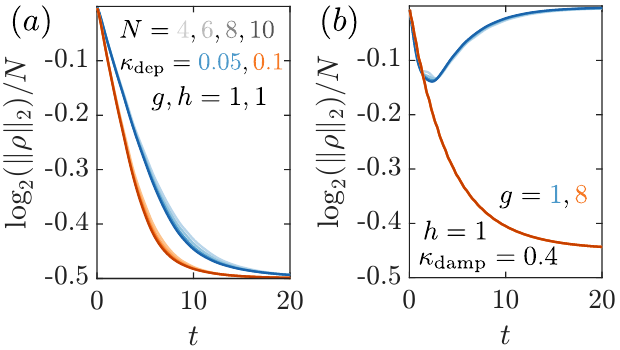}
        \caption{Evolution of the $L_2$ norm $\|\rho_t\|_2$ under Lindbladian dynamics of the non-integrable quantum Ising model with Ising interaction strength $J=1$ and transverse field $g$ and longitudinal field $h$. The continuous time $t$ is expressed in units of $J^{-1}$, with $\hbar=1$. The system sizes are $N=4,6,8,10$ (from lighter to darker colors). (a) Depolarizing noise with rate $\kappa_{\rm dep}=0.05, 0.1$, with $(g,h)=(1,1)$. (b) Same as (a),  but for amplitude-damping noise with rate $\kappa_{\rm damp}=0.4$, with $h=1$ and $g=1,8$.}
        \label{tfim_L2nm}
\end{figure}

The evolution of the $L_2$ norm of the density matrix, $\|\rho_t\|_2$, during the Lindbladian dynamics is shown in \cref{tfim_L2nm}. Similar to the case of noisy random circuits [cf.~\cref{l2nm_fig}], we observe that as the system size $N$ increases, $\|\rho_t\|_2=\sqrt{{\rm Tr}[\rho_t^2]}$ converges toward an asymptotic evolution 
\begin{equation} \label{L2_scale_eq_lind}
\frac{1}{N}\log_2(\|\rho_t\|_2)\approx-\lambda_p(t),
\end{equation}
where $\lambda_p(t)$ is system-size independent, but depends on the Hamiltonian parameters and on the noise type and strength. 
The system approaches its stationary purity on
an empirical timescale $T_s\sim1/\kappa$, with the steady-state $L_2$ norm characterized by $\lambda_p(t\gtrsim T_s)\approx  \lambda_p(\infty)$ [cf.~\cref{L2_scale_eq_lind}]. Owing to the data collapse in \cref{tfim_L2nm}, we use the numerical results for $N=10$ as a proxy for the behavior of $\lambda_p(t)$.

For the case of depolarizing noise [cf.~\cref{tfim_L2nm}(a)], we again see that $\|\rho_t\|_2$ monotonically reduces with time $t$ and evolves toward the maximally mixed state (corresponding to $\lambda_p^{\rm dep}(\infty)=0.5$). For amplitude-damping noise [cf.~\cref{tfim_L2nm}(b)], the $L_2$ norm initially decreases with time, may subsequently exhibit a revival, and eventually relaxes to a steady state distinct from the maximally mixed state, characterized by $\lambda_p^{\rm damp}(\infty)<0.5$. This is consistent with the same competition between coherent dynamics and the non-unital action of amplitude damping discussed in \cref{amp_damp_mech}.

\subsection{Noise-induced suppression of the $L_2$ error}

As in the noisy-circuit case [\cref{sec_l2_err}], we study the evolution of the single-step and cumulative $L_2$ errors incurred by MPO truncations during the Lindbladian dynamics. We consider the following setup in which the system density matrix is truncated only once, at a specific time $T_e$. This procedure produces an MPO denoted as $\sigma_{t,T_e}$, which evolves according to [cf.~\cref{lind_tro_channel}]
\begin{align} \label{L2_step_evo_lind}
\sigma_{t,T_e}=&\rho_t,\qquad \qquad \qquad \quad t<T_e \nonumber \\
\sigma_{t,T_e}=&({\cal T_R}\circ{\cal K}^{\cal L}_{\Delta t})(\sigma_{t-\Delta t,T_e}),\qquad t = T_e,\nonumber \\
\sigma_{t,T_e}=&{\cal K}^{\cal L}_{\Delta t}(\sigma_{t-\Delta t,T_e}), \qquad \qquad t > T_e.
\end{align}

\begin{figure}[t]
	\centering
	\includegraphics[width=0.48\textwidth]{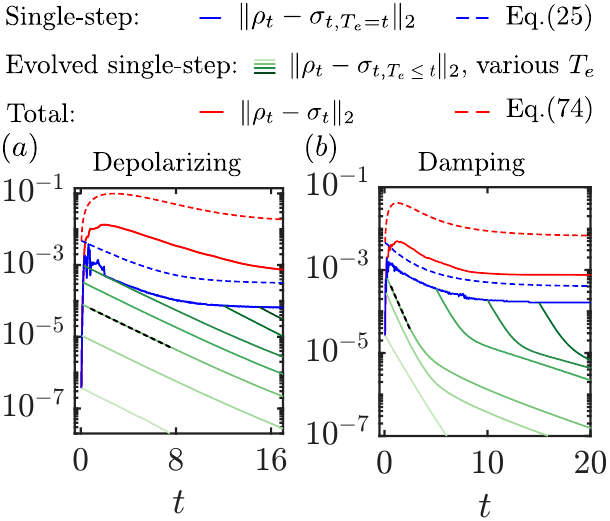}
        \caption{
        Behavior of the $L_2$ truncation error during Lindbladian evolution for (a) depolarizing noise, with $(g,h,\kappa_{\rm dep})=(1,1,0.05)$, and (b) amplitude-damping noise, with $(g,h,\kappa_{\rm damp})=(8,1,0.4)$. The continuous time $t$ is expressed in units of $J^{-1}$, with $J=1$. The system size $N=8$. The blue line denotes the single-step truncation error $\| \rho_t - \sigma_{t,T_e=t}\|_2$ [cf. \cref{L2_step_evo_lind}], and the blue dashed line denotes the corresponding empirical truncation error estimate \cref{trunc_error_tr}. The green lines (with multiple contrast levels) denote the evolution of the single-time truncation error $\| \rho_t - \sigma_{t,T_e\leq t}\|_2$ for several single-time truncation locations $T_e$. The red lines denote the total truncation error $\| \rho_t - \sigma_{t}\|_2$, and the red dashed lines denote the predicted empirical estimate \cref{L2scale_final_lind_full}. The dashed black segments indicate the data used to extract the suppression factor $\Gamma_p$ [\cref{err_shrink_lind_once}].}
        \label{tfim_L2err}
\end{figure}

In \cref{tfim_L2err}, we plot the single-step MPO truncation error $\| \rho_t - \sigma_{t,T_e=t} \|_2$ as a function of $t$ as the blue solid line, and the corresponding empirical estimate [\cref{trunc_error_tr}] as the blue dashed line. One observes that the initial truncation error is very small, followed by a short-time transient growth and oscillations. This indicates that the Lindbladian evolution requires some time to build up nontrivial entanglement structure from the initial product state, before the algorithm begins truncating singular values. After the initial evolution, we find good qualitative agreement between the single-step MPO truncation error and its corresponding empirical estimate [\cref{trunc_error_tr}].

The evolution of the single-step truncation error $\|\rho_t - \sigma_{t,T_e\leq t}\|_2$ for several different truncation locations $T_e$ is plotted as the green solid lines. We again observe the noise-induced suppression of this error, as for $t>T_e$:
\begin{equation} \label{err_shrink_lind_once}
\begin{aligned}
\left\|\rho_{t}-\sigma_{t,T_e}\right\|_2
&=\left\|{\cal K}_{\Delta t}^{\cal L}\left(\rho_{t- \Delta t}-\sigma_{t- \Delta t,T_e}\right)\right\|_2 \\
&\approx 2^{-\Gamma_p N \Delta t}\left\|\rho_{t- \Delta t}-\sigma_{t- \Delta t,T_e}\right\|_2,
\end{aligned}
\end{equation}
with a suppression factor $\Gamma_p>0$ that can be extracted from the data as indicated by the dashed black segments in \cref{tfim_L2err}. Compared with noisy random circuits, two differences arise here: (1) the decay rate of the single-step $L_2$ error $\|\rho_{t}-\sigma_{t,T_e}\|_2$ may differ from the decay rate of the $L_2$ norm [cf.~\cref{L2_scale_eq_lind}], 
and (2) under amplitude-damping noise, the single-step $L_2$ error can exhibit a two-stage decay with different suppression factors $\Gamma_p$ [cf.~\cref{tfim_L2err}(b)]. This reveals a slow mode in the Lindbladian evolution of the error matrix [cf.~\cref{err_shrink_lind_once}]. Although its small amplitude makes its contribution to the truncation error negligible, it dominates the long-time noise-induced suppression and is accounted for by our choice of $\Gamma_p$.

Our observed noise-induced suppression for the single-step truncation error [cf.~\cref{err_shrink_lind_once}] suggests that the Trotterized Lindbladian evolution channel suppresses the total truncation error as 
\begin{equation} \label{err_shrink_lind}
\left\|{\cal K}_{\Delta t}^{\cal L}\left(\rho_{t}-\sigma_{t}\right)\right\|_2 
\approx 2^{-\Gamma_p N \Delta t}\left\|\rho_{t}-\sigma_{t}\right\|_2.
\end{equation}

Following the derivation in \cref{app_L2_predict}, we estimate the total $L_2$ truncation error during the Lindbladian evolution as
\begin{widetext}
\begin{equation} \label{L2scale_final_lind_full}
\left\Vert \rho_{t}-\sigma_{t}\right\Vert _{2}\lesssim f(N,\delta_{\rm err})
    \sum_{j=1}^{t/\Delta t} 2^{-\Gamma_pN(t-j \Delta t)}
    \left\Vert {\cal K}^{\cal L}_{\Delta t}(\sigma_{(j-1)\Delta t}) \right\Vert _{2}.
\end{equation}
\end{widetext}

\begin{figure*}[t!]
	\centering
	\includegraphics[width=\textwidth]{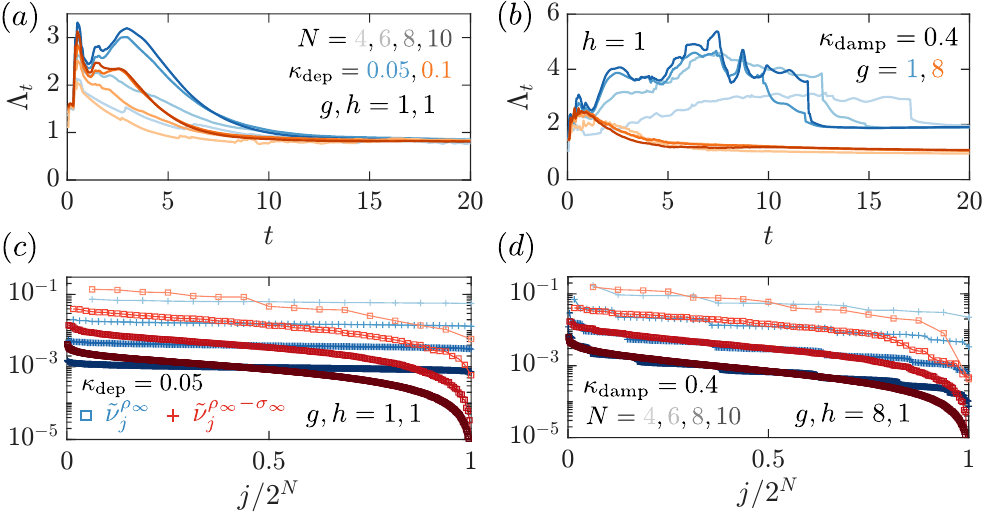}
        \caption{(a,b) Scaling behavior of the factor $\Lambda_t$ [\cref{lambda_t_defi_lind}] for Lindbladian dynamics, for the case of (a) depolarizing noise and (b) amplitude-damping noise. The parameters are the same as in \cref{tfim_L2nm}. (c,d) The normalized singular-value spectrum ${\tilde{\nu}_j^{\rho_\infty}}$ and ${\tilde{\nu}_j^{\rho_\infty-\sigma_\infty}}$ of the late-time state $\rho_\infty$ and of the corresponding late-time error matrix $\rho_\infty-\sigma_\infty$, for (c) depolarizing noise and (d) amplitude-damping noise. In all panels, the system sizes are $N=4,6,8,10$, from lighter to darker colors. }
        \label{TFIM_L1factor}
\end{figure*}

When the single-step compression error and
the accumulated relative $L_2$ error remain small,
we have $\left\Vert {\cal K}_{\Delta t}^{\cal L}(\sigma_{(j-1)\Delta t}) \right\Vert _{2} \approx \left\Vert \rho_{j\Delta t} \right\Vert _{2}$. Thus, by inserting the numerical behavior of $\left\Vert \rho_{j\Delta t} \right\Vert _{2}$ for all $j\in [1, t/\Delta t]$ [cf.~\cref{tfim_L2nm}] into \cref{L2scale_final_lind_full}, we obtain an empirical estimate for the $L_2$ error, shown as the red dashed lines in \cref{tfim_L2err}. The actual $L_2$ error incurred by the algorithm, $\left\Vert \rho_{t}-\sigma_{t} \right\Vert_{2}$, is shown in the same plot as the red solid line. We observe good qualitative agreement between the empirical estimate and the observed error evolution. The remaining quantitative discrepancy is due to the looseness of both the single-step empirical error prediction \cref{trunc_error_tr} and the triangle inequality, similar to \cref{tri_ineq}, as well as the overestimation of the error during the initial stage of the evolution, when the actual error remains very small because the state is close to a product state.

Finally, in the steady-state regime, we can further tighten the empirical estimate in \cref{L2scale_final_lind_full}. To this end, consider evolution times $t \gtrsim T_s + 1/\Gamma_p$, so that the system has reached its steady state, and the truncation error accumulated before steady state has damped away due to the noise-induced error suppression [cf.~\cref{err_shrink_lind}]. In this regime, the system has a fixed $L_2$ norm $\|\rho_\infty\|_2 \approx 2^{-\lambda_{p}(\infty)N}$ [cf.~\cref{L2_scale_eq_lind}]. Using the same derivation as in \cref{app_L2_predict}, we obtain the empirical steady-state estimate
\begin{equation} \label{L2scale_final_lind_steady}
\begin{aligned}
    \left\Vert \rho_{\infty}-\sigma_{\infty}\right\Vert _{2}&\lesssim f(N,\delta_{\rm err})
    \sum_{j=1}^{t/\Delta t} 2^{-\Gamma_pN\Delta t(t/\Delta t-j)}
    \cdot \|\rho_{\infty} \|_2\\
    &\lesssim 
     \frac{f(N,\delta_{\rm err})}{1-2^{-\Gamma_p N\Delta t}}\cdot  \|\rho_{\infty} \|_2,
\end{aligned}
\end{equation}
where we take the limit $t\to\infty$ to obtain the second line.

For $\Gamma_pN\Delta t\ll1$,
the denominator in \cref{L2scale_final_lind_steady}
scales as
$1-2^{-\Gamma_pN\Delta t}
\approx(\ln2)\Gamma_pN\Delta t$.
Thus, reducing $\Delta t$ at fixed compression
threshold can increase the estimated cumulative
truncation error. The accuracy requirement on
$\delta_{\rm err}$ must therefore account for
the Trotter step size.

\subsection{Evolution of the $L_1$ error, and a potentially efficient MPO algorithm for Lindbladian steady state}
\label{lind_l1_sec}

For the $L_1$ error, we again make use of the factor $\Lambda_t$ in \cref{lambda_t_defi} and formally write
\begin{equation} \label{lambda_t_defi_lind}
	 \left\Vert \rho_{t}-\sigma_{t}\right\Vert _{1} = \Lambda_t \cdot \frac{\left\Vert \rho_{t}-\sigma_{t}\right\Vert _{2}}{\left\Vert \rho_{t}\right\Vert _{2}}.
\end{equation}
Thus, both the behavior of $\Lambda_t$ and the ratio between the $L_2$ error and the $L_2$ norm determine the scaling of the $L_1$ error.

We plot the factor $\Lambda_t$ during the Lindbladian evolution for various system sizes $N$ in both the depolarizing and amplitude-damping cases in \cref{TFIM_L1factor}(a,b). As in the noisy random-circuit case [cf.~\cref{L1_rat_fig}], we observe nontrivial system-size dependence of $\Lambda_t$ before the system reaches its steady state. In the steady-state regime, however, $\Lambda_t$ again appears
approximately size independent over the accessible system sizes,
consistent with $\Lambda_\infty=O(1)$. Similarly to \cref{spectrum_sec}, we show in \cref{TFIM_L1factor}(c,d) the normalized singular-value spectrum of the steady state $\rho_\infty$ and of the corresponding error matrix $\rho_\infty-\sigma_\infty$. 
Once again, the spectra of $\rho_\infty$
and $\rho_\infty-\sigma_\infty$ have similar
shapes across the accessible system sizes,
consistent with the observed late-time scaling
$\Lambda_\infty=O(1)$ in
\cref{TFIM_L1factor}(a,b).

Therefore, by combining the steady-state empirical estimate for the ratio between the $L_2$ error and the $L_2$ norm in \cref{L2scale_final_lind_steady}, together with the steady-state convergence of $\Lambda_t$, we obtain an empirical estimate for the steady-state $L_1$ error:
\begin{equation} \label{L1scale_final_lind}
\begin{aligned}
    \left\Vert  \rho_{\infty}-\sigma_{\infty}\right\Vert _{1}
    \lesssim  \frac{f(N,\delta_{\rm err})}{1-2^{-\Gamma_p N\Delta t}} \Lambda_\infty. 
\end{aligned}
\end{equation}
Following a similar reasoning to that presented near \cref{circ_L1_tvd}, at fixed $\Delta t$ and noise strength, choosing $\delta_{\rm err}=O(\varepsilon^2/N^2)$ with a sufficiently small prefactor yields an $O(\varepsilon)$ empirical steady-state $L_1$ estimate. If output weights are signed, the conservative sampler bound in \cref{je_signed_sampling_bound} instead requires $\delta_{\rm err}=O(\varepsilon^2/N^4)$ for TVD $O(\varepsilon)$. We further provide numerical evidence that the operator entanglement of the MPO exhibits area-law scaling throughout the Lindbladian dynamics considered in this work [cf.~Appendix \ref{apd_area_sop}]. Taking these pieces of evidence together, we expect that the MPO simulation algorithm [cf.~\cref{mpo_lind_algo_sec}] may efficiently sample from the steady state of the Lindbladian dynamics studied in this work with small TVD.

We finally comment on a difference between the sampling algorithm for the Lindbladian case and that for the noisy random circuit case [cf.~\cref{mpo_circ_samp_algo}]. In both settings, the factor $\Lambda_t$ exhibits nontrivial growth with the system size $N$ before approaching its steady-state value [cf.~\cref{L1_rat_fig,TFIM_L1factor}]. For noisy random circuits, however, one may choose not to truncate the MPO in this regime, so that the algorithm remains efficient for arbitrary layers of circuits. 
For Trotterized Lindbladian dynamics, reaching
the steady-state regime requires approximately
$T_s/\Delta t$ time steps. When $\Delta t$ is
small, evolving exactly through this transient
regime can require prohibitively large MPO bond
dimensions. 
 Therefore, for Lindbladian dynamics, our current evidence supports only a potentially efficient MPO algorithm for sampling from the Lindbladian steady state.

\section{Outlook}
\label{sec_discuss}

Our work provides a systematic analysis of how MPO truncation errors evolve in 1D noisy random circuits and Lindbladian dynamics.  We reveal a previously overlooked phenomenon: noisy dynamics can exponentially suppress previously generated truncation errors, leading to substantially tighter empirical error estimates and, in the non-unital cases studied here, an exponential improvement over generic norm-based estimates.  A solvable global random-unitary model places the $L_2$ observations on firmer analytic footing: the same channel-dependent factor governs both the mean-squared contraction of traceless errors and the transient mean purity, while exact channel formulas reproduce the observed weak-noise behavior and the non-unital stationary-purity trend.  These results complement studies of area-law operator entanglement in 1D noisy random circuits~\cite{Noh2020,wei2025measurement} and 1D Lindbladian dynamics~\cite{PhysRevA.108.012616,PhysRevLett.129.170401} by addressing not only compressibility but also the evolution of approximation errors.  

There are several aspects where our results can be strengthened or extended:
\begin{itemize}

\item \textit{From the solvable global random-unitary model to adaptive MPO errors.} The global random-unitary model gives exact mean-squared contraction and mean-purity recurrences under fresh global scrambling.  The central remaining step is to determine when analogous contraction and concentration properties hold for the adaptively generated MPO errors in finite-depth local circuits and interacting Lindbladian dynamics. Analytically establishing such results would turn the present mechanism into a rigorous statement for the simulated dynamics.  Separately, an analytic explanation of the late-time singular-value-spectrum collapse and $\Lambda_\infty=O(1)$ would strengthen the resulting $L_1$ and TVD guarantees.

    \item \textit{Asymptotic scaling of $\Lambda_t$ before the steady state.} For both noisy random circuits and Lindbladian dynamics, we observe nontrivial growth of the factor $\Lambda_t$ (which controls the $L_1$ error) with system size $N$ during the transient regime. Since exactly computing the $L_1$ error requires exact diagonalization, our current numerics cannot access the asymptotic scaling of $\Lambda_t$ in this regime. It would therefore be interesting to develop alternative approaches to characterize the large-$N$ behavior of $\Lambda_t$, which could in turn lead to tighter error estimates for MPO-based simulation algorithms.
      
    \item \textit{Connection to other simulation algorithms.} Besides MPO-based methods, several other approaches can be used to simulate noisy quantum systems, including (1) quantum-trajectory methods~\cite{PRXQuantum.4.040326,cichy2025classical,sander2025large}, (2) Pauli-path truncation~\cite{PhysRevLett.133.230403,10.1145/3564246.3585234,schuster2024polynomial}, and (3) Clifford-based simulation~\cite{aaronson2004improved}. In these approaches, noise can facilitate classical simulation by (1) effectively unraveling the dynamics into measurements that suppress trajectory entanglement, or by (2,3) reducing the weight of long Pauli paths and the Clifford generators. It would be interesting to understand whether—and in what sense—these mechanisms can be related to the noise-induced suppression mechanism identified here.

 \item \textit{Extension to other settings.} We expect the noise-induced suppression of MPO truncation errors to be a general phenomenon across a broad range of noisy quantum dynamics, as already illustrated here for noisy random circuits and Lindbladian dynamics. It would therefore be interesting to explore such ``error dynamics'' in other settings, where it may likewise lead to tighter empirical error estimates. In particular, noisy boson sampling processes~\cite{PhysRevA.108.052604,PhysRevA.104.022407}, noisy monitored random circuit evolution (which is closely related to the sampling of two-dimensional noisy random circuits~\cite{wei2025measurement}), and non-Markovian noisy dynamics~\cite{Strathearn2018,PRXQuantum.3.010321} provide promising platforms for investigating these effects.

\item \textit{Noise-induced error suppression in other contexts.} Beyond the noise-induced suppression of MPO truncation errors studied in this work, recent results have shown that noise can also reduce errors in quantum simulation~\cite{xu2025exponentially} and yield tighter bounds on the energy under noisy time evolution~\cite{mishra2024classically}. Together with our results, these findings suggest that noise-induced error suppression may be broadly useful in both classical and quantum simulation contexts.
\end{itemize}

\textit{Note added.} While finalizing this manuscript, we became aware of Ref.~\cite{dowling2026classicalsimulabilityoperatorentanglement}, which provides sufficient conditions under which area-law scaling of operator entanglement implies that a Heisenberg-evolved operator admits an efficient MPO representation. This complements our study: we focus on the time evolution of truncation errors (in regimes where operator entanglement remains low) and identify a noise-induced suppression mechanism that helps maintain the accuracy and efficiency of MPO simulations of the density matrix.

\section*{Acknowledgements}
We thank Ignacio Cirac, Esther Cruz, and Bill Fefferman for insightful discussions.
ZYW, JR, and AVG acknowledge support from the U.S.~Department of Energy, Office of Science, Accelerated Research in Quantum Computing, Fundamental Algorithmic Research toward Quantum Utility (FAR-Qu). JR, JN and MJG acknowledge support from NSF QLCI (award No.~OMA-2120757) and NQVL:QSTD:Design:ORAQL. ZYW and AVG were also supported in part by NSF QLCI (award No.~OMA-2120757), NQVL:QSTD:Design:FTL, NSF STAQ program, DoE ASCR Quantum Testbed Pathfinder program (awards No.~DE-SC0019040 and No.~DE-SC0024220), ONR MURI,  AFOSR MURI,  DARPA SAVaNT ADVENT, and ARL (W911NF-24-2-0107). ZYW and AVG also acknowledge support from the U.S.~Department of Energy, Office of Science, National Quantum Information Science Research Centers, Quantum Systems Accelerator (award No.~DE-SCL0000121).
JN is supported by the National Science
Foundation Graduate Research Fellowship Program under Grant No. DGE 2236417.
DM acknowledges financial support from VILLUM FONDEN via the QMATH Centre of Excellence (Grant No.~10059) and from the Novo Nordisk Foundation under grant numbers NNF22OC0071934 and NNF20OC0059939.
JE has been supported by the BMFTR (MuniQC-Atoms, QuSol, Hybrid++, PasQuops, PraktiQom, Hybrid++), Clusters of Excellence (ML4Q, MATH+), the Munich Quantum Valley, Berlin Quantum, the Quantum Flagship (Millenion, Pasquans2), the DFG (CRC 183, SPP 2514), the QuantERA (SDPCode), and the European Research Council (DebuQC).
The numerical calculations
have been performed using ITensor.jl~\cite{itensor}.

The scientific content of the original arXiv version (v1) was developed entirely by the authors, with AI used only for language editing. For the revised version (v2), OpenAI's ChatGPT (GPT-6) was additionally used to assist in developing the solvable global Haar-random-unitary model presented in Sec. IV. The authors guided the analytical development, independently checked the resulting derivations, and take full responsibility for the scientific content.

\appendix

\section{Sharper normalization bounds}
\label{app:normalization}
Compared to \cref{err_bound_normalize_formal}, a sharper, fully a posteriori bound follows by removing the identity component before applying the triangle inequality. Define
\begin{equation}
\Xi_0=\Xi-\frac{\epsilon}{d}\mathbb I_d,
\qquad \rho_{\rm c}=\rho-\frac{\mathbb I_d}{d},
\end{equation}
so that both operators are traceless. Here $d=2^N$ is the Hilbert-space dimension. Hilbert--Schmidt orthogonality to the identity gives
\begin{equation}
\begin{aligned}
\|\rho-\mathcal T_{\mathcal R}(\rho)\|_2
&=\frac{\|\Xi_0-\epsilon\rho_{\rm c}\|_2}{|1-\epsilon|}\\
&\leq\frac{\sqrt{\|\Xi\|_2^2-|\epsilon|^2/d}}{|1-\epsilon|}\\
&\quad+\frac{|\epsilon|\sqrt{\|\rho\|_2^2-1/d}}{|1-\epsilon|}.
\end{aligned}
\label{je_centered_normalization}
\end{equation}
This estimate requires only $\epsilon\ne1$ and is no weaker than the corresponding uncentered triangle bound. If only the Schmidt-tail estimate is available, $\|\Xi\|_2^2$ in its numerator may be replaced by $2\sum_k\delta_k\|\rho\|_2^2$. In particular, the contribution proportional to $|\epsilon|$ is controlled by the distance from the maximally mixed operator, rather than the full purity. No dimension-independent control of $\epsilon$ follows from a small absolute residual alone: the general bound is $|\epsilon|\leq\sqrt d\,\|\Xi\|_2$.

The same decomposition suggests an alternative trace correction,
\begin{equation}
\begin{aligned}
\mathcal T_{\rm aff}(\rho)&=\widetilde\rho+\frac{\epsilon}{d}\mathbb I_d,\\
\|\rho-\mathcal T_{\rm aff}(\rho)\|_2^2
&=\|\Xi\|_2^2-\frac{|\epsilon|^2}{d},
\end{aligned}
\label{je_affine_trace_correction}
\end{equation}
which is the orthogonal projection of $\widetilde\rho$ onto the affine space of trace-one operators. It increases the MPO bond dimension by at most one and eliminates the normalization denominator entirely. Like scalar normalization, it does not ensure positivity. It is an analytically controlled alternative; the reported numerics use the original scalar normalization.

\section{Sampling from signed MPO weights}
\label{je_sampling_appendix}
In this appendix, we prove 
the validity of \cref{je_signed_sampling_bound}. Let $p$ be a probability distribution on $N$ bits and $q$ any real weights whose prefix sums can be evaluated. For a prefix $y$, set $a_b=p(yb)$, $v_b=q(yb)$, $u_b=[v_b]_+$, and $S_y=\sum_bu_b$. If $S_y>0$, the conditional rule in \cref{je_signed_sampler} satisfies
\begin{equation}
\begin{aligned}
\sum_b|a_b-p(y)Q(b|y)|
&\leq\sum_b|a_b-u_b|+|S_y-p(y)|\\
&\leq2\sum_b|a_b-v_b|.
\end{aligned}
\end{equation}
The last step uses $a_b\geq0$, so clipping a negative $v_b$ cannot increase its distance from $a_b$. If $S_y=0$, the same bound holds for any conditional distribution, since the left-hand side is at most $2p(y)$ and $\sum_b|a_b-v_b|\geq p(y)$. A telescoping comparison of the two sequential distributions, replacing one conditional at a time and averaging over prefixes drawn from $p$, yields
\begin{equation}
\begin{aligned}
\|p-Q\|_{\ell_1}
&\leq\sum_{k=1}^N\sum_{|y|=k-1}
p(y)\|p(\cdot|y)-Q(\cdot|y)\|_{\ell_1}\\
&\leq2\sum_{k=1}^N\sum_{|z|=k}|p(z)-q(z)|\\
&\leq2N\|p-q\|_{\ell_1}.
\end{aligned}
\end{equation}
Prefixes of zero $p$-weight contribute zero, and the last inequality follows from the triangle inequality for marginal sums. Dividing by two gives the TVD bound. For the Hermitian part of an MPO, the final comparison follows from the
contractivity of the trace norm under measurement and $\|\rho-\sigma_H\|_1\leq\|\rho-\sigma\|_1$. The argument assumes exact prefix contractions; numerical contraction and conditional-probability errors can be included as separate sampling errors.

\section{Details of the Lindblad dynamics}
\label{app:lindblad_trotter}

In this appendix we provide the explicit second-order (Strang) Trotterization
used in the main text for simulating Lindblad dynamics.
Throughout we split the generator into a coherent (Hamiltonian) part and local,
on-site dissipative terms, and we choose the single-qubit noise parameters such
that the discrete-time channel matches the intended continuous-time rates in the
$\Delta t \to 0$ limit. 

\subsection{Strang splitting between coherent and dissipative parts}

We start from the Lindblad master equation \cref{lind_equ} with Lindbladian $\mathcal L$, with the Hamiltonian part $\mathcal L_H$ and dissipator $\cal D$ specified in \cref{hami_equ,diss_equ,diss_form}.
For a time step $\Delta t$, the formal evolution is $\rho(t+\Delta t)=e^{\Delta t\mathcal L}\rho(t)$.
A second-order Strang splitting yields
\begin{equation}
e^{\Delta t\mathcal L}
=
\exp\!\Bigl(\frac{\Delta t}{2}\mathcal D\Bigr)\;
\exp(\Delta t\mathcal L_H)\;
\exp\!\Bigl(\frac{\Delta t}{2} \mathcal D\Bigr)
+O(\Delta t^3),
\label{eq:strang_split}
\end{equation}
where the $O(\Delta t^3)$ term denotes the
per-step Trotter error at fixed $N$.

\subsection{Second-order Trotterized unitary channel}
\label{subsec:brickwall_unitary}
For the coherent part, we decompose the Hamiltonian into nearest-neighbor terms
\begin{equation}
H=\sum_{i=1}^{N-1}H_{i,i+1},
\end{equation}
with the two-site operators within the bulk chosen as
\begin{equation}
H_{i,i+1}
= -J\,\tau_i^z\tau_{i+1}^z
-\frac{g}{2}\bigl(\tau_i^x+\tau_{i+1}^x\bigr)
-\frac{h}{2}\bigl(\tau_i^z+\tau_{i+1}^z\bigr).
\label{eq:two_site_H}
\end{equation}
The boundary terms are adjusted such that
$\sum_{i=1}^{N-1}H_{i,i+1}=H$.
Define the two-site unitary $U_{i,i+1}(\Delta t)=e^{-i\Delta t H_{i,i+1}}$ and its induced
channel
\begin{equation}
\mathcal U_{i,i+1}(\Delta t)(\rho)=U_{i,i+1}(\Delta t)\,\rho\,U_{i,i+1}^\dagger(\Delta t).
\end{equation}
Writing $\mathcal U^{\rm exact}_{\Delta t}(\rho)=e^{-i\Delta tH}\rho e^{i\Delta tH}$ for the exact unitary channel, we define its second-order approximation used in \cref{lind_tro_channel} by
\begin{equation}
\mathcal U_{\Delta t}:=
\prod_{i\in\mathrm{even}}\mathcal U_{i,i+1}(\tfrac{\Delta t}{2})
\prod_{j\in\mathrm{odd}}\mathcal U_{j,j+1}(\Delta t)
\prod_{k\in\mathrm{even}}\mathcal U_{k,k+1}(\tfrac{\Delta t}{2}),
\label{eq:brickwall_trotter}
\end{equation}
where ``even'' denotes bonds $(1,2),(3,4),\dots$ and ``odd'' denotes bonds
$(2,3),(4,5),\dots$.
This approximation satisfies $\mathcal U_{\Delta t}=\mathcal U^{\rm exact}_{\Delta t}+O(\Delta t^3)$ per step at fixed $N$.

\subsection{Implementation of the dissipators}
\label{diss_impl}
In our setting, the dissipator $\cal D$ [cf.~\cref{diss_equ}] consists of single-qubit dissipators that act independently on each site, so the dissipative
propagator over a duration $\Delta t$ factorizes into a product of identical
single-qubit quantum channels
\begin{equation}\label{diss_fator_apd}
\exp\!\Bigl(\Delta t\mathcal D\Bigr)
\;\equiv\;
\mathcal E_{\Delta t}^{\otimes N}.
\end{equation}
Each on-site CPTP map $\mathcal E_{\Delta t}$ is applied in Kraus form,
\begin{equation}\label{eq:kraus_general}
\begin{aligned}
&\mathcal E_{\Delta t}(\rho)=\sum_{\mu}K_{\mu}(\Delta t)\,\rho\,K_{\mu}^\dagger(\Delta t),\\
&\sum_{\mu}K_{\mu}^\dagger(\Delta t)K_{\mu}(\Delta t)=\mathbb I.
\end{aligned}
\end{equation}
We parameterize the channel strength through
$\mathcal E_{\Delta t}
=\exp(\Delta t\mathcal D_i)$,
where $\mathcal D_i$ is the single-site dissipator.
 Thus the Kraus operators are of the same form as those in \cref{kraus_dep,kraus_damp}, with corresponding noise probabilities
\begin{align}
    	p_{\rm dep}(\Delta t)&=\frac{3}{4}(1-e^{-4\kappa_{\rm dep} \Delta t/3}), \\
        p_{\rm damp}(\Delta t)&=1-e^{-\kappa_{\rm damp} \Delta t}.
\end{align}
Combining Eqs.~\eqref{eq:strang_split} and \eqref{eq:brickwall_trotter}, and implementing the dissipative propagators using the single-qubit maps in \cref{diss_fator_apd}, we obtain the per-step update in \cref{lind_evo,lind_tro_channel} in the main text.

\subsection{Scaling of operator entanglement in Lindbladian dynamics}
\label{apd_area_sop}
\begin{figure}[h!]

	\centering
	\includegraphics[width=0.4\textwidth]{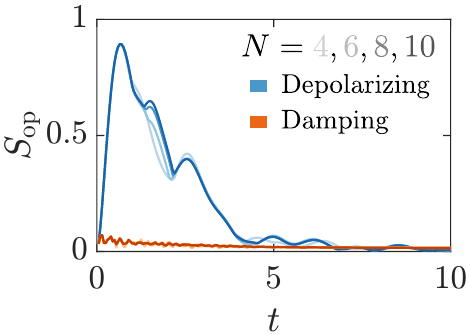}
        \caption{Scaling with system size $N$ of the operator entanglement $S_{\rm op}$ for the Lindbladian dynamics studied in the main text [\cref{lind_num_sec}]. The parameters match those in \cref{tfim_L2nm}: $(g,h,\kappa_{\rm dep})=(1,1,0.05)$ for depolarizing noise and $(g,h,\kappa_{\rm damp})=(8,1,0.4)$ for amplitude-damping noise.}
        \label{sop_fig}
\end{figure}

We present numerical evidence that, under the Lindbladian dynamics described in \cref{lind_l1_sec}, the \emph{operator entanglement} $S_{\rm op}$~\cite{PhysRevA.63.040304} of the density matrix $\rho_t$ across the half-chain bipartition obeys an area law. To this end, we define the vectorization of $\rho_t$ as $\ket{\rho_t}$, and the associated normalized ``density matrix'' for $\ket{\rho_t}$ as
\begin{equation}
\omega_t = \frac{|\rho_t\rangle \langle\rho_t| }{\operatorname{tr}[\rho_t^2]}.
\end{equation}
For a bipartition $AB$, the operator entanglement is the von Neumann entropy of the reduced state of $\omega_t$~\cite{PhysRevA.63.040304},
\begin{equation}\label{eq:Sop_def}
S_{\rm op}(t) := -\operatorname{tr}\big[\omega_t^{A}\log \omega_t^{A}\big] ,
\end{equation}
with $\omega_t^{A}= \tilde {\operatorname{tr}}_B [\omega_t]$, where $\tilde {\operatorname{tr}}$ denotes the partial trace of operators on the vectorized (doubled) Hilbert space.

In \cref{sop_fig}, we show the scaling of the half-chain operator entanglement $S_{\rm op}$ with system size $N$ for the Lindbladian dynamics studied in the main text [cf.~\cref{tfim_L2nm}]. Throughout the evolution, $S_{\rm op}$ is nearly independent of $N$, providing evidence of area-law scaling.

These von Neumann entropy results do not establish
the stronger uniform R\'enyi-$\alpha$ area-law
assumption with $0<\alpha<1$ required for
polynomial control of MPO bond dimensions.
 For depolarizing noise ($\kappa_{\rm dep}=0.05$), we observe a transient growth and subsequent decay of the operator entanglement due to the interplay between the Hamiltonian evolution and the depolarizing noise. For amplitude-damping noise, by contrast, $S_{\rm op}$ remains low throughout the evolution because we choose a relatively strong damping rate ($\kappa_{\rm damp}=0.4$) to accentuate the non-unital character of the noise, which also strongly suppresses entanglement growth.

\section{Additional details on the $L_2$ norm $\|\rho_t\|_2$}
\label{l2_amp_apd}
In the following, we discuss the behavior of the $L_2$ norm in more detail, focusing on the case of noisy random circuit evolution [cf.~\cref{sec_norm}]. In \cref{pure_noise_sol,depo_sp_apd}, we provide further evidence for the scaling form in \cref{L2_scale_eq} by comparing its asymptotic regimes with exact purely noise-driven dynamics, and by numerically characterizing the late-stage deviation from the steady-state $L_2$ norm for the case of depolarizing noise. Finally, in \cref{amp_damp_mech}, we give a physical
interpretation of the stationary late-time purity under
amplitude-damping noise and connect it to the global random-unitary
result in \cref{haar_purity_sec}. We expect the same mechanism to apply to Lindbladian evolution under amplitude-damping noise as well [cf.~\cref{tfim_L2nm}(b)].

\subsection{Dynamics under pure noise}\label{pure_noise_sol}
In \cref{l2nm_fig}, we saw that the behavior of the $L_2$ norm $\|\rho_t\|_2$ approximately collapses onto the form in \cref{L2_scale_eq}. To provide further evidence, we consider the same initial product-state
setting as in \cref{circ_set}, but without random
unitary gates. Here $t$ counts individual
applications of the product-noise channel
$\mathcal N=\phi^{\otimes N}$, whereas each full
circuit layer in the main text contains two
noise applications. In this case, the system density matrix remains a product state throughout the evolution, and exact expressions for $\|\rho_t\|_2$ are readily obtained.

\textit{Purely depolarizing dynamics.} In this case, starting from an arbitrary initial pure product state, the $L_2$ norm evolves as
\begin{equation} \label{pure_dep_l2}
\frac{1}{N}\log_2 (\|\rho_t\|_2)
=
\frac{1}{2} \log_2\left(
\frac{1+\left(1-\frac{4}{3}p_{\rm dep}\right)^{2t}}{2}
\right).
\end{equation}
In the regime of weak fixed noise, $p_{\rm dep}\ll 1$ (which is the more interesting and difficult regime for classical simulation), \cref{pure_dep_l2} simplifies in the short-time regime $t\ll 1/p_{\rm dep}$ to
\begin{equation} \label{apd_st_dep}
\frac{1}{N}\log_2 (\|\rho_t\|_2)\approx -\frac{2}{3\ln2}p_{\rm dep}\, t.
\end{equation}
At late times and weak noise, the leading correction gives
\begin{equation} \label{apd_lt_dep}
	\frac{1}{N}\log_2 (\|\rho_t\|_2)\approx -\frac{1}{2} + \frac{1}{2\ln2}e^{-(8/3)p_{\rm dep}t}.
\end{equation}
This pure-noise result exhibits the same qualitative
behavior as \cref{L2_scale_eq}: an initial exponential
decay of $\|\rho_t\|_2$, with a rate
$2p_{\rm dep}/(3\ln 2)$ per noise application
[cf.~\cref{apd_st_dep}], followed by relaxation
toward the maximally mixed state, characterized by
$\lambda_p=1/2$, with an exponentially decaying
correction to the logarithmic norm at the weak-noise
rate $8p_{\rm dep}/3$ [cf.~\cref{apd_lt_dep}].

\begin{figure}[t!]
	\centering
	\includegraphics[width=0.48\textwidth]{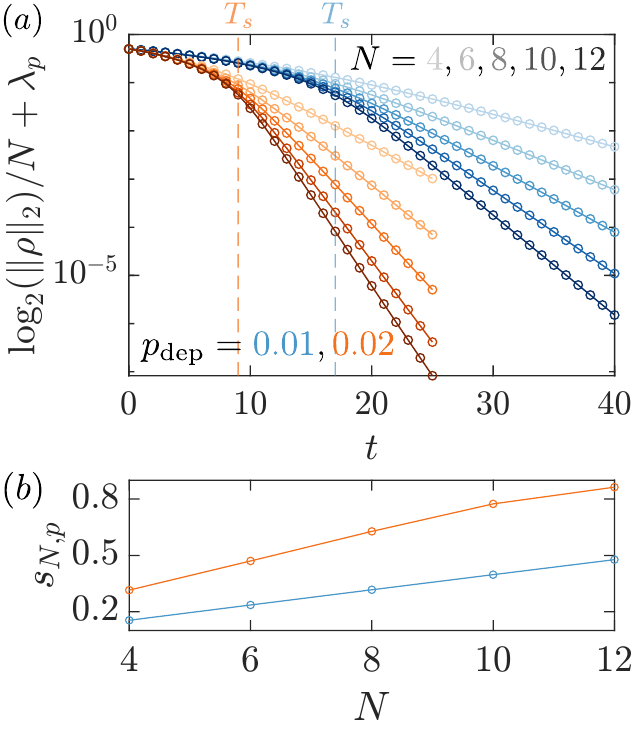}
        \caption{(a) Evolution of the deviation from the steady-state $L_2$ norm $\|\rho_t\|_2$ in noisy circuits, plotted as $\frac{1}{N}\log_2 (\|\rho_t\|_2)+\lambda_p$, under depolarizing noise with rates $p_{\rm dep}=0.01,0.02$, for system sizes $N=4,6,8,10,12$ (from lighter to darker colors). The number of circuit layers $t$ takes integer values. The dashed lines indicate the corresponding timescales $T_s$. (b) The extracted decay slope $s_{N,p_{\rm dep}}$ [cf.~\cref{late_apd_decay_form}] from the late-stage behavior shown in panel (a).}
        \label{st_devi_fig}
\end{figure}

\textit{Purely amplitude-damping dynamics.}
In this case, we consider the initial state
$\rho_0=|1,\cdots,1\rangle\langle1,\cdots,1|$,
which undergoes nontrivial relaxation under
amplitude damping. The $L_2$ norm then evolves as
\begin{align}\label{l2_damp_apd}
&\frac{1}{N}\log_2(\|\rho_t\|_2)\nonumber\\
&=\frac{1}{2}\log_2\left(
\left[1-(1-p_{\rm damp})^t\right]^2
+(1-p_{\rm damp})^{2t}
\right).
\end{align}
For fixed weak noise $p_{\rm damp}\ll1$,
\cref{l2_damp_apd} simplifies in the short-time
regime $t\ll1/p_{\rm damp}$ to
\begin{equation}\label{damp_st_apd}
\frac{1}{N}\log_2(\|\rho_t\|_2)
\approx-\frac{p_{\rm damp}}{\ln2}\,t.
\end{equation}
At late times, the leading correction is
$-(1-p_{\rm damp})^t/\ln2$, which for weak noise
gives
\begin{equation}\label{damp_lt_apd}
\frac{1}{N}\log_2(\|\rho_t\|_2)
\approx-\frac{1}{\ln2}e^{-p_{\rm damp}t}.
\end{equation}
Unlike pure depolarizing noise, amplitude damping
produces a nonmonotonic purity evolution: the
initially pure excited state first becomes mixed
and eventually relaxes to the pure ground state.
Random unitary layers continually regenerate
excitations, leading to the nontrivial stationary
mixed-state behavior observed in \cref{l2nm_fig}.

\subsection{Late-time relaxation of the logarithmic
$L_2$ norm [cf.~\cref{L2_scale_eq}]} \label{depo_sp_apd}
Here we numerically characterize the late-time relaxation of the logarithmic $L_2$ norm, focusing on noisy random-circuit evolution under depolarizing noise, since in this case the exact steady-state $L_2$ norm, characterized by $\lambda_p=0.5$, can be determined exactly. 

Specifically, we replot the data in \cref{l2nm_fig}(a) in terms of the deviation from the steady-state value, $\frac{1}{N}\log_2 (\|\rho_t\|_2)+\lambda_p$, as shown in \cref{st_devi_fig}. One can then clearly see the two-stage behavior separated by the timescale $T_s$, with the deviation in the late stage, $t\gtrsim T_s$, exhibiting an exponential decay of the form
\begin{equation} \label{late_apd_decay_form}
	\frac{1}{N}\log_2 (\|\rho_t\|_2)+\lambda_p \sim e^{-s_{N,p_{\rm dep}}t},
\end{equation}
with an exponent $s_{N,p_{\rm dep}}$ that depends on both the system size $N$ and the noise rate $p_{\rm dep}$, as extracted in \cref{st_devi_fig}(b).
The observed relaxation rate $s_{N,p_{\rm dep}}$ increases with $N$, indicating that the late-time correction decays increasingly rapidly. We therefore neglect this correction in the empirical truncation-error estimates in the main text [cf.~\cref{L2_scale_eq}].

\subsection{Physical interpretation of the stationary purity under amplitude damping}
\label{amp_damp_mech}

The stationary late-time purity under amplitude-damping
noise can be understood from the competition between
scrambling and the non-unital action of the channel.
Amplitude damping drives each qubit toward $\lvert0\rangle$,
which tends to purify the state, while also suppressing
coherences.  The unitary dynamics continually scrambles the
state and regenerates coherences, so the stationary purity is
set by the balance between these effects.

At weak damping, the non-unital bias toward
$\lvert0\rangle$ is small, and scrambling produces a highly
mixed stationary regime, corresponding to
$\lambda_p\approx1/2$.  At strong damping, the reset tendency
dominates and the state remains closer to
$\lvert0\cdots0\rangle$, so $\lambda_p$ decreases toward
zero.  This explains qualitatively the monotonic trend in
\cref{l2nm_fig}.

The global random-unitary model in \cref{haar_purity_sec} makes this
many-body picture analytically tractable.  For product
amplitude damping, it yields the exact finite-$N$ stationary
mean purity and the large-$N$ exponent in
\cref{haar_damp_purity_largeN}.
It therefore reproduces the same monotonic trend while also
connecting the purity dynamics to the contraction of
traceless errors.

\bibliography{library.bib}
\end{document}